\documentclass{article} % For LaTeX2e
\usepackage{./style} 
\usepackage{times}
\usepackage[utf8]{inputenc}
\usepackage[T1]{fontenc}

\usepackage{amsmath,amsfonts,bm}

\def\eqref#1{equation~\ref{#1}}
\def\1{\bm{1}}

\DeclareMathAlphabet{\mathsfit}{\encodingdefault}{\sfdefault}{m}{sl}
\SetMathAlphabet{\mathsfit}{bold}{\encodingdefault}{\sfdefault}{bx}{n}

\usepackage{hyperref}
\usepackage{xurl}

\usepackage{booktabs}       % professional-quality tables
\usepackage{amsfonts}       % blackboard math symbols
\usepackage{nicefrac}       % compact symbols for 1\2, etc.
\usepackage{microtype}      % microtypography
\usepackage{xcolor}         % colors
\usepackage{multirow}
\usepackage{booktabs}
\usepackage{amsmath}
\usepackage{amssymb}
\usepackage{mathtools}
\usepackage{amsthm}
\usepackage{algorithm}
\usepackage{algorithmic}
\usepackage{enumitem}
\usepackage{natbib}
\usepackage{caption}
\usepackage{newunicodechar}
\usepackage{fancyvrb,fvextra}
\usepackage{tikz-cd}
\usepackage{listings} 
\usepackage{makecell}
\usepackage{textcomp} 
\usepackage{framed}
\usepackage{graphicx}
\usepackage{subcaption}
\usepackage{cleveref} 
\usepackage{enumitem}
\usepackage{float}
\usepackage[most]{tcolorbox}

\usepackage{color}
\definecolor{keywordcolor}{rgb}{0.7, 0.1, 0.1}   % red
\definecolor{tacticcolor}{rgb}{0.0, 0.1, 0.6}    % blue
\definecolor{commentcolor}{rgb}{0.4, 0.4, 0.4}   % grey
\definecolor{symbolcolor}{rgb}{0.0, 0.1, 0.6}    % blue
\definecolor{sortcolor}{rgb}{0.1, 0.5, 0.1}      % green
\definecolor{attributecolor}{rgb}{0.7, 0.1, 0.1} % red

\tcbuselibrary{listings,breakable}
\newtcolorbox[auto counter,number within=section]{lstlistingbox}[2][]{%
    breakable,
    enhanced,
    colback=gray!5,
    colframe=gray!50,
    boxrule=0.5pt,
    arc=2pt,
    left=6pt, right=6pt, top=6pt, bottom=6pt,
    fonttitle=\bfseries,
    title={Listing \thetcbcounter: #2},
    #1
}

\usepackage{courier}  % Monospace font (replaces FreeMono)
\usepackage{helvet}   % Sans-serif font (replaces DejaVu Sans)
\renewcommand{\sfdefault}{phv}  % Use Helvetica as the default sans-serif font

\newunicodechar{ζ}{{\ensuremath{\zeta}}}

\title{LeanCat: A Benchmark Suite for Formal Category Theory in Lean (Part I: 1-Categories)}

\author{\textbf{Rongge Xu}$^{1}$\thanks{Email: \href{mailto:rongge@tsinghua.edu.cn}{\tt rongge@tsinghua.edu.cn}} \quad \textbf{Hui Dai}$^{2}$ \quad \textbf{Yiming Fu}$^{3}$ \quad \textbf{Jiedong Jiang}$^{4}$ \quad \textbf{Tianjiao Nie}$^{5}$ \quad \textbf{Junkai Wang}$^1$\quad\!
\\\textbf{Holiverse Yang}$^6$\quad\!\textbf{Zhi-Hao Zhang}$^7$\\
$^1$Yau Mathematical Sciences Center, Tsinghua University \quad\!$^2$Iluvatar CoreX \quad\! $^3$Department\\
of Mathematics, Southern University of Science and Technology \quad\! $^4$Westlake Institute for \\
 Advanced Study, Westlake University \quad\!$^5$ Qiuzhen College, Tsinghua University \\ $^6$Department of Physics, The Chinese University of Hong  Kong \quad\! $^7$Yanqi Lake Beijing Institute \\ of Mathematical Sciences and Applications (BIMSA)\\
}

\iclrfinalcopy

\begin{document}

\maketitle

\begin{abstract}
While large language models (LLMs) have demonstrated impressive capabilities in formal theorem proving, current benchmarks fail to adequately measure library-grounded abstraction---the ability to reason with high-level interfaces and reusable structures central to modern mathematics and software engineering. We introduce \textbf{LeanCat}, a challenging benchmark comprising 100 fully formalized category-theory tasks in Lean. Unlike algebra or arithmetic, category theory serves as a rigorous stress test for structural, interface-level reasoning. Our evaluation reveals a severe \textit{abstraction gap}: the best state-of-the-art model solves only $12.0\%$ of tasks at pass@4, with performance collapsing from $55.0\%$ on Easy tasks to $0.0\%$ on High-difficulty tasks, highlighting a failure in compositional generalization. To overcome this, we evaluate \textbf{LeanBridge}, a retrieval-augmented agent that employs a \textit{retrieve--generate--verify} loop. LeanBridge achieves a peak success rate of $24.0\%$---doubling the performance of the best static baseline. These results empirically demonstrate that iterative refinement and dynamic library retrieval are not merely optimizations but strict necessities for neuro-symbolic reasoning in abstract domains. LeanCat offers a compact, reusable testbed for tracking progress toward reliable, research-level formalization.
\end{abstract}

\section{Introduction}\label{sec:intro}

Recent advances in large language models (LLMs) and agentic training have revived the prospect of \emph{end-to-end} formal theorem proving.
On the formal side, systems such as OpenAI's early Lean-based prover \citep{Openai2022} and DeepMind's AlphaProof \citep{DeepMind2025} demonstrate that reinforcement learning with formal verification feedback can produce non-trivial Lean proofs.
More recently, specialized provers such as DeepSeek-Prover-V2 \citep{deepseekproverv2} and Seed-Prover 1.5 \citep{SeepProver1.5} show that large-scale agentic RL and test-time scaling can push formal success rates on established benchmarks substantially higher.
These results suggest that formal proof generation is no longer limited to toy domains, and that tight tool feedback loops (retrieve--generate--verify) can be a decisive ingredient.

\begin{figure*}[t]
    \centering
    \includegraphics[width=0.7\textwidth]{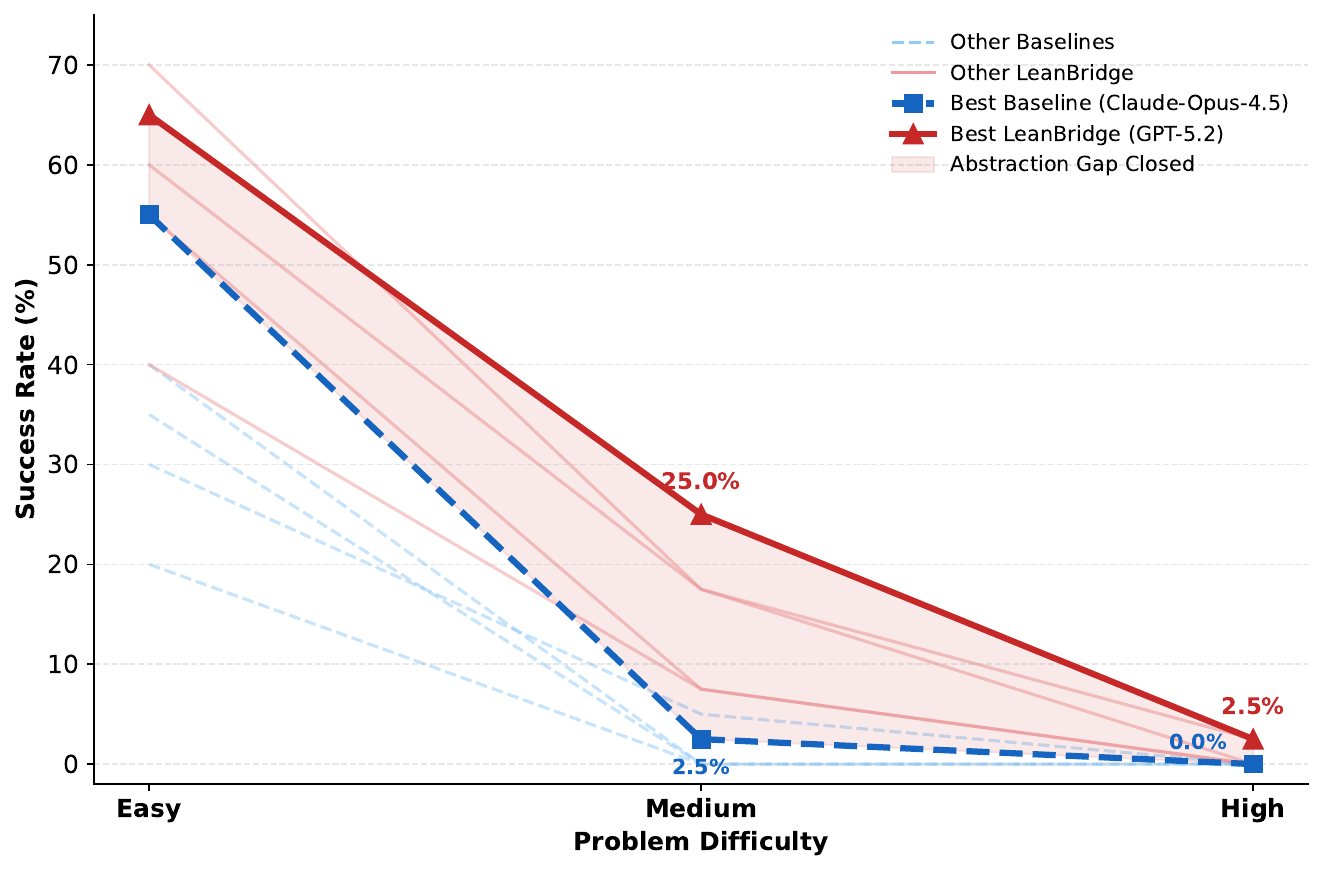}
    \caption{\textbf{Closing the Abstraction Gap}. 
    Performance comparison of standard baselines (dashed lines) vs. our agentic approach, LeanBridge (solid lines), across difficulty levels. 
    \textbf{Bold lines} highlight the best-performing models (Claude-Opus-4.5 for pass@4 baseline, GPT-5.2 for LeanBridge with \emph{NL+Statement} input), while faint lines represent other models. 
    Note the sharp performance collapse of baselines on Medium/High tasks (the ``abstraction gap''), which LeanBridge effectively bridges, achieving the first successful proofs on High-difficulty problems.}
    \label{Formalization_accuracy}
\end{figure*}

Despite steady progress in neural theorem proving, current formal benchmarks still do not adequately exercise \emph{library-grounded, abstraction-heavy} reasoning. Widely used datasets such as miniF2F \citep{miniF2F} and FIMO \cite{FIMO} largely draw from olympiad-style problems, while university-level suites like ProofNet \cite{proofNet} and PutnamBench \cite{putnamBench} focus on undergraduate competition or textbook material. These benchmarks are valuable, but they often reward short clever tricks or computation rather than sustained work inside rich abstract frameworks. By contrast, modern research mathematics is written at high generality, organized around reusable interfaces, and deeply intertwined with large libraries of definitions and lemmas—so success hinges less on a single key insight and more on navigating abstraction, managing definitions, and coherently composing library knowledge over long horizons.

Category theory provides a natural stress-test for this capability: as the interface language of modern mathematics—categories, functors, natural transformations, adjunctions, limits/colimits, monads—formal proofs typically rely on \emph{diagrammatic} and \emph{universal-property} reasoning, i.e., constructing morphisms with the right naturality/uniqueness guarantees and showing commutativity across families of structures. Yet existing formal benchmarks rarely target this abstraction layer directly. To bridge this gap, we introduce \textbf{LeanCat}\footnote{The benchmark code and data are available at \url{https://github.com/sciencraft/LeanCat}.}, a benchmark of 100 category-theory problems formalized in Lean 4 (with Mathlib), designed to test whether automated provers can operate \emph{inside} a mature library and compose high-level abstractions rather than solve isolated puzzles. LeanCat complements algebra-focused benchmarks such as the FATE series \cite{Fate} by shifting the frontier from abstract algebra to category theory. 

Our baseline evaluation reveals a stark \textbf{abstraction gap}: across five strong models, the best achieves only \textbf{$8.25\%$} at pass@1 and \textbf{$12\%$} at pass@4. More critically, we observe a collapse in compositional generalization: performance drops from $55.0\%$ on Easy tasks to \textbf{0.0\%} on High-difficulty tasks (Figure~\ref{Formalization_accuracy}), validating that current LLMs struggle to manage long-horizon abstraction dependencies.

To address this challenge, we go beyond static evaluation and introduce \textbf{LeanBridge}, a retrieval-augmented agent designed to close the abstraction gap. LeanBridge implements a dynamic \textit{retrieve--generate--verify} loop that actively queries the library for definitions and lemmas. Our experiments show that LeanBridge achieves a success rate of $24.0\%$. As shown in Figure~\ref{Formalization_accuracy}, our agentic approach (solid lines) effectively mitigates the performance collapse observed in baselines, securing the first successful proofs on High-difficulty tasks. These results empirically demonstrate that iterative refinement and library retrieval are not merely optimizations, but strict necessities for neuro-symbolic reasoning in abstract domains.

To our knowledge, LeanCat is the first installment of a benchmark series dedicated to category theory. Beyond merely ranking models, we view this direction as pivotal for both \emph{human mathematics}---by mapping which parts of abstract reasoning remain elusive to formalization---and \emph{AI}---by forcing progress on abstraction-aware planning, definition retrieval, and robust compilation-driven refinement.

Our key contributions are summarized as follows:
\begin{itemize}
    \item \textbf{LeanCat benchmark (1-Categories).}
    We present LeanCat, a suite of 100 category-theory problems formalized in Lean 4. The tasks span eight topic clusters, curated to cover reusable abstract interfaces and structural reasoning rather than contest-style tricks.
    \item \textbf{Difficulty annotation pipeline.}
    We propose a hybrid grading process combining model-based estimates and expert judgment. Each problem receives weighted ratings from both advanced LLMs and human formalizers to assign reliable Easy/Medium/High labels.
    \item \textbf{Baseline evaluation.}
    We benchmark state-of-the-art generalist models (e.g., GPT-5.2, Claude-Opus-4.5) alongside specialized neural provers (e.g., DeepSeek-Prover-V2). Our extensive evaluation reveals a severe \textit{abstraction gap}: while generalist models struggle with long-horizon coherence, specialized provers fail to generalize beyond their training curriculum, achieving near-zero success on Medium/High tasks even with high sampling budgets.
    \item \textbf{Search-augmented proving via LeanBridge.}
    To address the abstraction gap, we propose and evaluate \textbf{LeanBridge}, an agentic workflow integrating LeanExplore \citep{Asher_LeanExplore_2025}. We demonstrate that this retrieval-augmented approach doubles the pass rate of static baselines, validating the necessity of dynamic tool use in higher mathematics.
\end{itemize}

\section{LeanCat Benchmark Design}\label{sec:methodology}

\subsection{Benchmark Structure and Content}\label{subsec:benchmark-curation}

\paragraph{Data Source}
The problems in our benchmark are organized into two main parts: Abstract and Concrete:
    \begin{itemize}
    \item \textbf{Abstract part}: Problems are primarily drawn from standard, widely used textbooks in category theory, especially Category Theory in Context \citep{Riehl2017} and Categories for the Working Mathematician \citep{ML1998}, together with a small number of problems adapted from unpublished lecture notes \citep{Kong, Zheng}.
    \item \textbf{Concrete part}: Problems are curated mainly from Abstract and Concrete Categories \citep{AHS1990}, which provides a systematic source of exercises on concretizability, injectivity, and related themes.
    \item \textbf{Others}: Beyond these core sources, we also include selected problems inspired by research papers and advanced community-driven references \citep{Chen2021,ACMU2021}.
    \end{itemize}

Each LeanCat problem is self-contained at the statement level: the formal statement of the theorem is provided (often with informal description provided in the corresponding Markdown file), and all necessary definitions exist in Lean's environment (either already in Mathlib \citep{mathlib} or introduced as part of the problem setup). 
Where possible, we drew on known theorems from category theory literature, many tasks were crafted or adapted specifically to probe corner cases and interface interactions that an AI prover might struggle with.
In several High-difficulty cases, relevant lemmas were not readily available, forcing human formalizers to devise intermediate results. This aspect of LeanCat, that it sometimes goes beyond the library, makes it a particularly stringent test for automated provers, which cannot rely solely on rote application of existing library facts.

LeanCat comprises 100 theorem statements in category theory, each fully formalized in Lean 4 (i.e. each problem is given as a Lean theorem declaration, with required definitions and context available). The problems are organized into eight topical clusters reflecting key areas of category theory, ranging from Basic Category Properties to advanced structures like Abelian Categories and Monads.

\begin{figure}[htbp]
        \centering
        \includegraphics[scale=0.2]{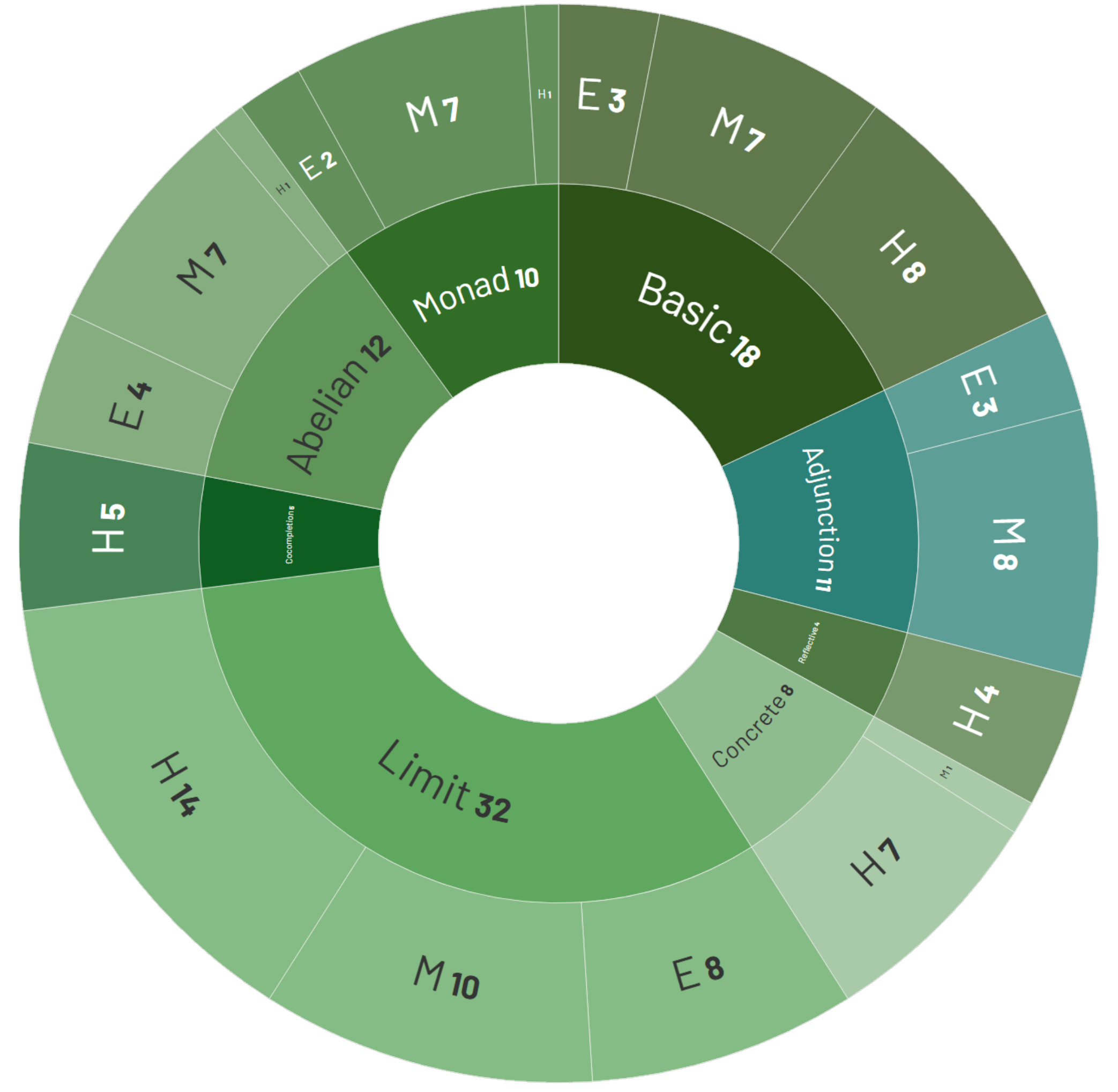}
        \caption{Sunburst diagram showing the distribution of our LeanCat problem sets by topic and difficulty. The inner ring groups problems into thematic sections (Basic, Adjunction, Reflective, Concrete, Limit, Cocompletion, Abelian, Monad), while the outer ring lists individual problems, labelled by difficulty: E = Easy, M = Medium, H = High.}
        \label{sunburst_cat}
\end{figure}

Figure \ref{sunburst_cat} illustrates the distribution of problems across these topics and their difficulty levels. The benchmark covers both standard textbook results (e.g., adjunctions, limits) and frontier formalization tasks (e.g., cocompletions, which require new definitions). We provide a detailed breakdown of each topical cluster and the specific mathematical competencies they test in Appendix \ref{app:topic_details}.

\subsection{Curation Workflow}
LeanCat is curated through a three-stage workflow that combines expert selection, LLM-assisted drafting, and rigorous human verification:
\begin{enumerate}
    \item \textbf{Collection.}
    Three category-theory experts curate candidate problems from established sources (as described above), aiming to cover both core interfaces (e.g., adjunctions, limits/colimits, monads) and representative proof patterns (diagram chasing, universal properties, naturality).
    \item \textbf{Formalization.}
    For each selected problem, we first use several LLMs to draft Lean 4 statements.
    The same three category-theory experts then review these drafts and retain only semantically correct formal statements.
    For problems where none of the models produced a correct statement, we organized a three-day workshop at Westlake University, bringing together Lean experts to author the missing statements and (when feasible) corresponding proofs.
    \item \textbf{Review.}
    Finally, two independent reviewers with strong mathematical backgrounds and Lean expertise perform a full consistency pass, checking compilation, fixing definitional mismatches, and ensuring the formal statement matches the intended mathematical meaning.
\end{enumerate}

\paragraph{Statement-level tasks.}
LeanCat is a \emph{statement-level} benchmark: each item consists of a single standalone theorem to prove (see Figure~\ref{fig:sample_task} in Appendix \ref{app:topic_details} for a concrete example), rather than a scaffolded sequence of intermediate lemmas that gradually builds toward a final goal.
This design aims to evaluate general library-grounded proving capabilities---retrieval, definition management, and abstraction navigation---instead of rewarding problem-specific hint engineering.

\paragraph{Scope and difficulty.}
Overall, LeanCat is \emph{broad} in its topical coverage across category theory, and \emph{deep} in the sense that even short-looking theorems can require layered abstractions and careful use of reusable interfaces, mirroring how mathematicians work inside large formal libraries.

\paragraph{Formalization standards.}
All benchmark files follow strict, uniform conventions:
(i) each Lean file is strictly self-contained, relying solely on Mathlib and explicitly including any definitions or lemmas required by the theorem statement; (ii) the natural-language problem description is stored in a companion Markdown file (supporting \LaTeX{}), designed to be injected into the Lean code as a comment during the evaluation phase.

\subsection{Difficulty Annotation Pipeline}\label{subsec:benchmark-characteristics}
    
Rather than relying purely on the problem author's intuition, we implemented a systematic LLM+human rating pipeline to grade problem difficulty on a 10-point scale, then binned the scores into the three categories: Easy, Medium, and High. 
This approach aimed to capture both human expertise and automated solver perspectives, similar in spirit to FATE's curation process (which combined expert judgment with model feedback for difficulty ranking).

Our pipeline proceeded as follows:

\begin{itemize}
    \item \textbf{LLM Difficulty Scoring.} We employ a data-driven approach where each problem is evaluated using the LeanBridge sequential refinement protocol (as detailed in Section \ref{sec:experiments}) across all five baseline models. We assign 2 points for a full proof and 0.5 points for a correct statement (credited only if the proof fails). The difficulty is calculated as: $\text{Diff} = \max(0, 10 - 2 N_{\text{proof}} - 0.5 N_{\text{stmt}})$, where $N_{\text{proof}}$ is the number of models producing a verified proof, and $N_{\text{stmt}}$ is the number of models generating a correct formal statement but failing the proof (ensuring disjoint counts). This scales from 0 (all solved) to 10 (none solved).
    % We employ a data-driven approach where problems are attempted by multiple baseline models. Scores are derived based on whether models can generate correct statements or proofs. Problems solved by many models receive lower scores, while those unsolved by any model receive higher scores. To obtain a robust difficulty estimate, we evaluate each problem using the \textbf{LeanBridge sequential refinement protocol} (as detailed in Section \ref{sec:experiments}) across all five baseline models. The difficulty score is calculated as follows:
    %     \begin{equation}
    %         \text{Diff} = \max(0, 10 - \text{PF score} - \text{ST score})
    %     \end{equation}
    %     where:
    %     \begin{itemize}
    %         \item \textbf{PF score} (Proof Score): The contribution from models that successfully generated a complete, compiling proof.
    %         \item \textbf{ST score} (Statement Score): The contribution from models that generated a correct formal statement (but failed the proof). This is only counted if the model did not already contribute to the PF score.
    %     \end{itemize}
    %     This mechanism ensures that problems solved by none of the models receive a difficulty of 10, while problems solved by all models approach 0.
    \item \textbf{Human Difficulty Scoring.} In parallel, two human mathematicians (with Lean expertise and category theory background) independently rated each problem on the same 1-10 difficulty scale. 
    They factored in things like the length of the proof, intricacy of arguments, and whether any non-obvious lemmas are required. The human scores tended to correlate with intuition: e.g. a trivial diagram chase might be 2/10, whereas a complex construction spanning several definitions could be 9/10.
    \item \textbf{Aggregation.} We combined the scores giving 50\% weight to human ratings and 50\% to LLM ratings. Finally, we mapped the numeric scores to difficulty categories by selecting thresholds specifically to enforce a distribution of \textbf{20 Easy, 40 Medium, and 40 High} tasks. This resulted in cut-offs of scores $\le 5.54$ for Easy and $\ge 7.8$ for High. We intentionally adopted this skewed distribution (heavier on Medium/High tasks) to prevent premature saturation by current models---which already solve $>50\%$ of Easy tasks---and to provide sufficient headroom for tracking future progress in closing the abstraction gap. The complete scoring data for all 100 problems is provided in Appendix Figure~\ref{app:fig:score_sample}.
\end{itemize}
            
This data-driven approach provides a more objective difficulty metric than human intuition alone. Our experimental results (Section 4) validate this ranking: the best-performing model solved most tasks classified as \textbf{Easy}, while problems with scores $\ge 9$ remained unsolved across all baselines. This distribution confirms that our difficulty labels effectively proxy the "formalization hardness" for current neuro-symbolic systems.

\section{Experiments and Results}\label{sec:ex_and_results}

\subsection{Experimental Setup}\label{sec:experiments}

We employ two distinct inference strategies to evaluate performance on LeanCat: parallel sampling for standard baselines, and sequential refinement for the LeanBridge agent.

\paragraph{Standard Baselines: Parallel Sampling (Pass@k).} We benchmark a suite of state-of-the-art models, including GPT-5.2 \citep{ChatGPT5.2}, Gemini-3-Pro \citep{Gemini3pro}, Claude-Opus-4.5 \citep{ClaudeOpus4.5}, DeepSeek-V3.2-Thinking \citep{DeepseekV3.2}, and Kimi-K2 \cite{KimiK2}, under a purely formal setup. 
\begin{itemize} 
    \item \textbf{Input:} Each problem is presented in a uniform format containing the Lean statement (theorem name, hypotheses, and conclusion) and necessary imports. The model sees the formal goal exactly as a human user would. Crucially, \emph{no natural language hints or decomposed lemmas are provided}; the model must derive the proof solely from the formal statement and standard library knowledge. 
    \item \textbf{Prompting:} We prompt models to generate a Lean proof script directly, adopting the prompt template from FATE-Eval \citep{Fate} for consistency. 
    \item \textbf{Protocol:} We use the \emph{pass@k} metric \citep{pass-k}. For each problem, we sample $k$ independent proof attempts. The problem is considered solved if \emph{any} of the $k$ attempts compiles and verifies. We report pass@1 (single-try reliability) and pass@4 (sampling diversity).
\end{itemize}

To determine if the abstraction gap is merely a result of using general-purpose models, we extended our pass@k evaluation to specialized neural provers explicitly trained for formal mathematics: DeepSeek-Prover-V2-671B\citep{deepseekproverv2}, Goedel-Prover-v2-32B\citep{goedel}, StepFun-Prover-32B\citep{stepfunprover2025}, and Kimina-72B\citep{kimina}. Given their specialized nature, we relaxed the evaluation strictness, allowing for a significantly higher sampling budget of pass@32 (compared to pass@4 for generalist models).

\begin{algorithm}[tb]
   \caption{LeanBridge Inference Loop (Dynamic Retrieval)}
   \label{alg:leanbridge}
\begin{algorithmic}
   \STATE {\bfseries Input:} Natural Language Statement $S_{NL}$, Formal Statement $S_{Lean}$ (Optional)
   \STATE {\bfseries Output:} Proof Script $\pi$ or Failure
   
   \STATE \textcolor{gray}{// Phase 1: Context Initialization \& Mandatory Retrieval}
   \STATE $C \leftarrow \{S_{NL}, S_{Lean}\}$
   \STATE $D_{init} \leftarrow \text{Retrieve}(Query=S_{NL})$ \COMMENT{Initial broad search}
   \STATE $C \leftarrow C \cup \{D_{init}\}$

   \STATE \textcolor{gray}{// Phase 2: Refinement Loop}
   \FOR{$t=1$ {\bfseries to} $T_{max}$}
       \STATE $\text{Action}_t \leftarrow \text{LLM}.\text{Generate}(C)$
       
       \IF{$t > 1$ \AND $\text{Action}_t$ is \textbf{SearchRequest}($\text{Keywords}$)}
           \STATE \textcolor{gray}{// Agent decides to refine knowledge}
           \STATE $D_{new} \leftarrow \text{Retrieve}(\text{Keywords})$
           \STATE $C \leftarrow C \cup \{\text{Action}_t, D_{new}\}$
           
       \ELSIF{$\text{Action}_t$ is \textbf{ProofCandidate}($\pi$)}
           \STATE \textcolor{gray}{// Agent attempts a proof}
           \STATE $Success, Msg \leftarrow \text{LeanCompiler}.\text{Verify}(\pi)$
           
           \IF{$Success$}
               \STATE \textbf{return} $\pi$
           \ELSE
               \STATE \textcolor{gray}{// Feedback loop: Error message drives next step}
               \STATE $C \leftarrow C \cup \{\pi, Msg\}$ 
           \ENDIF
       \ENDIF
   \ENDFOR
   \STATE \textbf{return} Failure
\end{algorithmic}
\end{algorithm}

\paragraph{LeanBridge: Sequential Refinement.} To address the library-grounded and abstraction-heavy nature of LeanCat, we also evaluate LeanBridge, a retrieval-augmented agent. Unlike the static baselines, LeanBridge actively leverages Mathlib via LeanExplore~\citep{Asher_LeanExplore_2025}---which utilizes a hybrid ranking strategy combining semantic embeddings, BM25+, and PageRank---through a \emph{retrieve--generate--verify} loop (Algorithm~\ref{alg:leanbridge}). 
\begin{itemize} 
    \item \textbf{Input:} We test two settings: (1) \emph{NL-to-Proof}: The model receives only the natural-language problem description; and (2) \emph{NL+Statement-to-Proof}: The model receives both the natural language description and the formal statement. 
    \item \textbf{Protocol:} Unlike the independent sampling used for baselines, LeanBridge operates in a sequential loop. The system initiates a search using the natural-language statement, generates a proof, and validates it. If verification fails, the system analyzes the error, potentially performs a targeted retrieval, and revises the proof. We allow a maximum of 4 iterations per problem. A problem is considered solved if the system produces a valid, verified proof at any point within these 4 iterations. 
\end{itemize}

\begin{table*}[t]
    \centering
    \caption{\textbf{Baseline Performance.} Success rates of static models evaluated via Pass@1 and Pass@4 metrics. Pass@1 scores are estimated using the unbiased estimator from \citet{Chen2021}.}
    \label{tab:model_formal_performance}
    \setlength{\tabcolsep}{5pt} % Adjust column spacing
    \begin{tabular}{lcccc}
    \toprule
    \textbf{Model} & \textbf{Easy} & \textbf{Medium} & \textbf{High} & \textbf{Overall}\\
    \midrule
    Claude-Opus-4.5
    & \textbf{40.00}\%~/~\textbf{55.00}\%
    & 0.63\%~/~2.50\%
    & 0.00\%~/~0.00\%
    & \textbf{8.25}\%~/~\textbf{12.00}\%\\
    
    GPT-5.2
    & 27.50\%~/~35.00\%
    & 0.00\%~/~0.00\%
    & 0.00\%~/~0.00\%
    & 5.50\%~/~7.00\%\\
    
    Gemini-3-Pro
    & 13.75\%~/~30.00\%
    & \textbf{1.25}\%~/~\textbf{5.00}\%
    & 0.00\%~/~0.00\%
    & 3.25\%~/~8.00\%\\
    
    DeepSeek-V3.2
    & 20.00\%~/~40.00\%
    & 0.00\%~/~0.00\%
    & 0.00\%~/~0.00\%
    & 4.00\%~/~8.00\%\\
    
    Kimi-K2
    & 10.00\%~/~20.00\%
    & 0.00\%~/~0.00\%
    & 0.00\%~/~0.00\%
    & 2.00\%~/~4.00\%\\
    \bottomrule
    \end{tabular}
\end{table*}

\begin{table*}[t]
  \centering
  \caption{\textbf{LeanBridge Performance.} Success rates of the agentic workflow (max 4 iterations). Columns distinguish between \textit{NL-only} and \textit{NL+Statement} input settings.}
  \label{tab:leanbridge_performance}
  \setlength{\tabcolsep}{5pt}
  \begin{tabular}{lcccc}
    \toprule
    \textbf{Model} & \textbf{Easy} & \textbf{Medium} & \textbf{High} & \textbf{Overall}\\
    \midrule
    Claude-Opus-4.5
    & \textbf{70.0}\%~/~\textbf{70.0}\%
    & 5.0\%~/~17.5\%
    & 0.0\%~/~0.0\%
    & \textbf{16.0}\%~/~21.0\%\\[2pt]
    
    GPT-5.2
    & 60.0\%~/~65.0\%
    & 5.0\%~/~\textbf{25.0}\%
    & 0.0\%~/~\textbf{2.5}\%
    & 14.0\%~/~\textbf{24.0}\%\\[2pt]
    
    Gemini-3-Pro
    & 45.0\%~/~60.0\%
    & \textbf{12.5}\%~/~17.5\%
    & 0.0\%~/~\textbf{2.5}\%
    & 14.0\%~/~20.0\%\\[2pt]
    
    DeepSeek-V3.2
    & 55.0\%~/~40.0\%
    & 0.0\%~/~7.5\%
    & 0.0\%~/~0.0\%
    & 11.0\%~/~11.0\%\\[2pt]
    
    Kimi-K2
    & 30.0\%~/~55.0\%
    & 2.5\%~/~7.5\%
    & 0.0\%~/~0.0\%
    & 7.0\%~/~14.0\%\\
    \bottomrule
  \end{tabular}
\end{table*}

\paragraph{Verification and Constraints.} All proof attempts are verified against Lean 4 (Mathlib v4.19.0). 
\begin{itemize} 
    \item \textbf{Success Criteria:} A proof is considered valid if the Lean kernel accepts it without errors and strictly contains no \texttt{sorry}, \texttt{admit}, or additional \texttt{axiom} declarations. 
    \item \textbf{Resource Limits:} Each attempt is subject to a generic output budget of 50,000 tokens and a verification time limit of 5 minutes. Attempts exceeding these limits are marked as failures. 
    \item \textbf{Human Review (NL-to-Proof Only):} In the \emph{NL-to-Proof} setting, a model may generate a valid proof for an incorrect formalization. To guard against such superficial ``green'' proofs, all successful candidates in this mode undergo human expert review to ensure semantic alignment with the original natural-language problem. 
\end{itemize}

\section{Results and Analysis}\label{Results}

We evaluate five state-of-the-art models on LeanCat. We first present the quantitative performance of standard baselines, followed by the results of our retrieval-augmented agent, LeanBridge. Finally, we provide a qualitative analysis of common failure modes.

\subsection{Quantitative Performance}

\paragraph{Standard Baselines (Pass@k).}
Table~\ref{tab:model_formal_performance} summarizes the success rates of API-based models in the purely formal setting.
As shown, the overall success rate is low across the board. The best-performing model, Claude-Opus-4.5, solves only \textbf{8.25\%} of problems on the first attempt (Pass@1) and improves to \textbf{12.00\%} with four attempts (Pass@4).
Other top-tier models like GPT-5.2 and DeepSeek-V3.2 follow, achieving Pass@4 scores of 7.00\% and 8.00\%, respectively.
Across all baseline experiments, only \textbf{14 distinct problems} were solved by at least one model, highlighting the rigorous difficulty of LeanCat compared to earlier benchmarks.

\begin{table}[t]
    \centering
    \caption{\textbf{Specialized Provers (Pass@32).} Performance of domain-specific models. Despite a higher sampling budget, they fail to generalize to Medium/High tasks.}
    \label{tab:specialized_provers}
    % 1. 缩小列间距
    \setlength{\tabcolsep}{3.5pt}
    % 2. 强制适应单栏宽度
    % \resizebox{\columnwidth}{!}{%
    \begin{tabular}{lcccc}
    \toprule
    \textbf{Model} & \textbf{Easy} & \textbf{Med.} & \textbf{High} & \textbf{Avg.} \\
    \midrule
    DeepSeek-Prover-V2-671B
    & \textbf{45.0}\% & 0.0\% & 0.0\% & \textbf{9.0}\%\\
    
    Goedel-Prover-V2-32B
    & 20.0\% & \textbf{2.5}\% & 0.0\% & 5.0\%\\
    
    StepFun-Prover-32B
    & 20.0\% & \textbf{2.5}\% & 0.0\% & 5.0\%\\
    
    Kimina-72B
    & 10.0\% & 0.0\% & 0.0\% & 2.0\%\\
    \bottomrule
    \end{tabular}%
    % }
\end{table}

% \begin{table*}[t]
%     \centering
%     \caption{Performance of Specialized Neural Provers (Pass@32). Unlike generalist baselines, these models were sampled 32 times per problem due to their specialized nature. Despite the higher compute budget, the ``abstraction gap'' remains severe.}
%     \label{tab:specialized_provers}
%     \setlength{\tabcolsep}{5pt}
%     \begin{tabular}{lcccc}
%     \toprule
%     \textbf{Model} & \textbf{Easy} & \textbf{Medium} & \textbf{High} & \textbf{Overall}\\
%     \midrule
%     DeepSeek-Prover-V2-671B
%     & \textbf{45.00}\%
%     & 0.00\%
%     & 0.00\%
%     & \textbf{9.00}\%\\
    
%     Goedel-Prover-V2-32B
%     & 20.00\%
%     & \textbf{2.50}\%
%     & 0.00\%
%     & 5.00\%\\
    
%     StepFun-Prover-32B
%     & 20.00\%
%     & \textbf{2.50}\%
%     & 0.00\%
%     & 5.00\%\\
    
%     Kimina-72B
%     & 10.00\%
%     & 0.00\%
%     & 0.00\%
%     & 2.00\%\\
%     \bottomrule
%     \end{tabular}
% \end{table*}

\paragraph{LeanBridge Performance.}
Table~\ref{tab:leanbridge_performance} details the performance of our agentic approach. By comparing static baselines, the NL-only agent, and the NL+Statement agent, we isolate two key drivers of success:

\begin{itemize}
    \item \textbf{The Agentic Advantage (Static vs. Agent).} 
    First, the retrieval loop alone provides a massive boost. Even without the formal statement (\textit{NL-only}), LeanBridge (GPT-5.2) achieves an overall success rate of \textbf{$14.0\%$}, doubling its static Pass@4 baseline of $7.0\%$. This confirms that dynamic library interaction effectively bridges basic knowledge gaps.

    \item \textbf{Impact of Formal Grounding (NL-only vs. NL+Statement).} 
    Second, comparing the two agent settings reveals that grounding the agent in the formal statement is decisive for handling \textbf{abstraction}. 
    While the two settings perform similarly on Easy tasks ($60.0\%$ vs. $65.0\%$), the gap widens drastically on Medium tasks ($5.0\%$ vs. \textbf{$25.0\%$}). 
    Crucially, only the \textit{NL+Statement} setting unlocks \textbf{High-difficulty} problems ($2.5\%$), a frontier that remained completely inaccessible ($0\%$) to both static baselines and ungrounded agents. 
    This finding indicates that precise formal definitions act as essential anchors for long-horizon planning in abstract domains. (See Appendix~Figure~\ref{app:fig:ablation_detailed} for a detailed visual comparison across all models.)
\end{itemize}

\paragraph{Comparison with Specialized Provers.}
Table \ref{tab:specialized_provers} presents the results. Three key observations emerge:
\begin{itemize}
    \item \textbf{The Abstraction Gap Persists.} Even with a $8\times$ higher sampling budget, specialized provers exhibit the same collapse in performance as generalist models. DeepSeek-Prover-V2, the strongest in this category, achieves a respectable $45.0\%$ on Easy tasks but plummets to $0.0\%$ on both Medium and High tasks.
    \item \textbf{Curriculum Overfitting.} The disparity between Easy and Medium/High tasks is even more pronounced in specialized provers than in generalist LLMs. For instance, Goedel-Prover achieves 20\% on Easy tasks but fails almost completely on Medium tasks (2.5\%). This suggests that these models may have overfitted to the ``olympiad-style'' or ``undergraduate calculation'' distribution prevalent in their training data (e.g., miniF2F), failing to generalize to the structural reasoning required for category theory.\item \textbf{Inefficiency of Blind Scaling.} Comparing Table \ref{tab:leanbridge_performance} and Table \ref{tab:specialized_provers}, we see that a generalist agent (LeanBridge with GPT-5.2), using a maximum of only 4 iterations, surpasses specialized provers at pass@32 on overall metrics, and significantly outperforms them on harder tasks. This indicates that agentic workflow and library grounding offer superior scaling potential regarding reasoning depth for abstract mathematics compared to simply scaling up sampling on model weights trained for lower-level proofs.
\end{itemize}

\subsection{Performance Analysis}

\paragraph{Difficulty Gap.}
As visualized in Figure~\ref{Formalization_accuracy}, there is a divergent trend between static baselines and our agentic framework.
For standard models, we observe a sharp performance degradation consistent with our difficulty annotations.
For instance, Claude-Opus-4.5 achieves \textbf{55.00\%} Pass@4 on Easy problems but drops to \textbf{2.5\%} on Medium and \textbf{0\%} on High.
In contrast, LeanBridge (solid lines in Figure~\ref{Formalization_accuracy}) mitigates this collapse, maintaining a smoother degradation curve and securing the only successes on High-difficulty tasks.
Notably, even the ``Easy'' subset---which requires nontrivial abstraction and library navigation---is far from solved. This suggests that the baseline complexity of LeanCat is already beyond the reliable capability of current models, which struggle once problems move beyond simple exercises.

\paragraph{Efficacy of Retries (Pass@1 vs. Pass@4).}
Moving from Pass@1 to Pass@4 yields modest absolute gains. For top models like Claude-Opus-4.5 / GPT-5.2, the improvement is marginal ($+3.75\%$ / $+1.5\%$). This saturation indicates that simple test-time scaling (more samples) is insufficient for structural reasoning; the models fail not because of random noise, but due to fundamental deficits in knowledge and planning. This observation motivates the necessity of the active retrieval mechanism in LeanBridge. 
% This indicates high variance in weaker models (where success depends on sampling a lucky strategy), whereas stronger models tend to fail deterministically on hard problems due to fundamental reasoning or knowledge gaps, rather than simple syntax errors that could be fixed by resampling.

\subsection{Qualitative Analysis and Failure Modes}

To understand the root causes of failure, we manually inspected failed attempts (specifically from the Medium and High categories). For a deep dive into specific examples, we refer readers to Appendix~\ref{app:case_study}, which present step-by-step trajectories of a representative success (Problem 59) and a complex failure (Problem 63). Based on this inspection, we identify five dominant failure modes:

\begin{enumerate}
  \item \textbf{Math failure.} The proof uses an incorrect mathematical strategy (e.g.\ vacuity for an existential goal, constant diagrams for genuine colimits), so no Lean patch can make it true. \emph{Example: in \textbf{Prob.~68} it picks an empty category \texttt{PEmpty} and tries to finish by \texttt{simp}, but the existential statement then reduces to \(\bot\)}.

  \item \textbf{Lean grammar failure.} The mathematical idea may be fine, but the Lean script is syntactically/elaborationally invalid (bad braces, wrong binder order, illegal notation, private constructors), so the verifier fails before reaching the proof content. \emph{Example: in \textbf{Prob.~73} it triggers parser/elaboration errors (e.g.\ ``unexpected token \texttt{have}; expected \texttt{\}}'') while attempting to build lifted-limit data}.

  \item \textbf{Hallucination failure.} The model invokes nonexistent lemmas/constants or misuses projections (e.g.\ \texttt{T.foo} as a field when it is not), often due to name guessing or version/import mismatch. \emph{Example: in \textbf{Prob.~3} it repeatedly uses nonexistent API such as \texttt{hFull.preimage} / \texttt{Functor.image\_preimage} to lift maps from \texttt{Full}, yielding ``unknown constant'' errors}.

  \item \textbf{Lazy failure.} The model avoids developing required intermediate definitions/lemmas not in mathlib and instead tries \texttt{infer\_instance}, guessed theorem names, comments, or placeholders, so the missing bridge is never constructed. \emph{Example: in \textbf{Prob.~75} it attempts \texttt{by classical infer\_instance} to prove \(\texttt{IsFiltered (Rec C)}\), and when that fails it outputs only commentary rather than proving the needed lemma}.

  \item \textbf{Hack failure.} The model makes the goal trivially provable by altering the formal meaning of the statement (e.g.\ redefining/shadowing key notions) rather than proving the intended theorem. \emph{Example: in \textbf{Prob.~17} it defines a local \texttt{Retract} that merely stores \(\texttt{IsInjectiveObj Y}\) and a \texttt{piObj} that ignores its input, so the theorem reduces to a tautology and compiles without proving the real claim}.
\end{enumerate}

Pure syntax-level errors (e.g., indentation or parenthesis mismatch) are present but significantly less frequent than the semantic and strategic failures described above.
 
% These failure modes reflect the inherent difficulty profile of \textbf{LeanCat}. Some problems are conceptually advanced and demand nontrivial categorical infrastructure rather than routine tactic search. Others hinge on constructing explicit examples or counterexamples with careful universe and instance management. A further class requires bridging highly abstract formulations (universal properties, presentability, completion conditions) with concrete encodings (explicit objects, diagrams, or algebraic presentations), where missing library glue or small definitional mismatches can derail the entire proof.

\begin{table}[t]
    \centering
    \caption{\textbf{Ablation Study (Gemini-3-Pro).} Comparisons of Baseline vs. LeanBridge variants. \textbf{Full (w/ Search)} achieves the best performance on harder tasks.}
    \label{tab:ablation_retrieval}
    % \resizebox{\columnwidth}{!}{%
        \begin{tabular}{lcccc}
        \toprule
        \multirow{2}{*}{\textbf{Diff.}} & \multicolumn{2}{c}{\textbf{Baselines}} & \multicolumn{2}{c}{\textbf{LeanBridge}} \\
        \cmidrule(lr){2-3} \cmidrule(lr){4-5}
         & \textbf{Pass@1} & \textbf{Pass@4} & \textbf{No Search} & \textbf{Search} \\
        \midrule
        \textbf{Easy}   & 13.75\% & 30.00\% & 45.00\% & \textbf{60.00}\% \\
        \textbf{Medium} & 1.25\%  & 5.00\%  & 7.50\%  & \textbf{17.50}\% \\
        \textbf{High}   & 0.00\%  & 0.00\%  & 0.00\%  & \textbf{2.50}\% \\
        \midrule
        \textbf{Avg.} & 3.25\% & 8.00\% & 12.00\% & \textbf{20.00}\% \\
        \bottomrule
        \end{tabular}%
    % }
\end{table}

\subsection{Ablation Study: The Impact of Retrieval}
To disentangle the contributions of the iterative agentic loop versus the dynamic library retrieval, we conducted an ablation study using Gemini-3-Pro. We evaluated a ``No-Search'' variant of LeanBridge, which retains the \textit{verify--refine} loop but is restricted to its internal parametric knowledge (i.e., forbidden from issuing search queries).

Table~\ref{tab:ablation_retrieval} presents the results. We observe a clear hierarchy of capability:
\begin{itemize}
    \item \textbf{Loop vs. Sampling:} LeanBridge (w/o Search) achieves $12.0\%$, significantly outperforming the static Pass@4 baseline ($8.0\%$). This indicates that the ability to interpret compiler error messages and self-correct provides a tangible gain over blind resampling, particularly on Easy tasks ($30.0\% \to 45.0\%$).
    \item \textbf{The Necessity of Retrieval:} However, the ``No-Search'' agent hits a ceiling on harder problems, solving only $7.5\%$ of Medium tasks and failing completely on High tasks ($0\%$). Enabling retrieval (Full LeanBridge) unlocks a further $\sim2\times$ performance jump on Medium tasks ($17.5\%$) and enables the only successes on High tasks. This confirms that while the agentic loop handles logical bugs, dynamic retrieval is indispensable for bridging the knowledge gap in abstract formalization.
\end{itemize}

LeanBridge substantially strengthens an LLM's practical proving ability, and retrieval is a key driver of that improvement. With retrieval enabled, the system more reliably escapes ``local patching'' plateaus by pulling in the right library knowledge at the right time, translating into markedly better performance on harder problems.

\section{Discussion and Future Work}\label{sec:discussion}
\textbf{Key Challenges in Abstract Formalization.}
Our results highlight four persistent bottlenecks that distinguish LeanCat from earlier benchmarks: (i) \emph{Library Awareness:} The primary hurdle is not logical reasoning per se, but locating and instantiating the correct abstract interfaces from Mathlib. (ii) \emph{Abstraction Control:} Models struggle to maintain the ``categorical imperative,'' often drifting into counter-productive element-wise reasoning that disconnects from the intended library lemmas. (iii) \emph{Long-horizon Coherence:} Maintaining a consistent proof plan across dependent steps remains difficult without external guidance or backtracking. (iv) \emph{The Limits of Specialization:} Our evaluation confirms that domain-specific fine-tuning is insufficient for LeanCat. Even with extensive sampling (pass@32), specialized provers (e.g., DeepSeek-Prover, Goedel) fail on Medium/High tasks. This suggests that abstract formalization requires the broad ``mathematical world knowledge'' and compositional reasoning found in large-scale generalist models, rather than merely local tactic prediction.

\textbf{The LeanCat Roadmap: Towards Higher Algebra.}
We envision LeanCat not merely as a static dataset, but as the first installment of a broader series targeting high-abstraction mathematics. Future iterations will expand beyond 1-category theory to cover monoidal categories and higher-categorical structures (e.g., 2-categories). This evolution aims to create a virtuous cycle:
% $$ \text{Benchmark} \rightarrow \text{AI Solutions} \rightarrow \text{Merged into Mathlib} \rightarrow \text{New, Harder Frontier} $$
\begin{equation*}
\begin{tikzcd}[row sep=small, column sep=small, ampersand replacement=\&]
    \text{Benchmark} \arrow[r] 
    \& \text{AI Solutions} \arrow[d] \\
    \text{New Frontier} \arrow[u, dashed] 
    \& \text{Merged to Mathlib} \arrow[l]
\end{tikzcd}
\end{equation*}

\textbf{Broader Impact.}
For \emph{AI research}, LeanCat provides a concrete testbed for improving retrieval-augmented generation (RAG) and abstraction-aware planning. For the \emph{mathematics community}, these checkpoints help identify missing lemmas and interface gaps in the library, prioritizing human engineering efforts.

% \subsection{Limitations}\label{sec:limitations}
% We acknowledge a primary limitation regarding our sampling budget. Our \emph{pass@k} evaluation uses relatively low sample counts ($k=1, 4$). While sufficient to expose the severe difficulty gap, this may not fully capture the tail of model distributions. However, our preliminary analysis indicates that the abstraction errors in LeanCat are structural rather than stochastic, implying that simple resampling is unlikely to yield significant gains without improved reasoning strategies.

\subsubsection*{Acknowledgments}

The Lean workshop at Westlake University was supported by the National Natural Science Foundation of China (NSFC), Grant No. 12574176. We thank Prof. Jian Li for supporting the organization of the workshop. We are also grateful to Jiatong Yang and Hongwei Wang for their valuable assistance with formalization during the workshop. We specifically thank Zhimin Cao for conducting the evaluation of specialized provers. We thank Xingyu Ren for inspiring this benchmark. We thank Prof. Lifan Guan, Huayi Chen, Bin Dong and Hao Zheng for helpful comments. We also thank Jiadong Adam Guo, Jinyuan Zhang, and Xiaodong Hu for providing API access for our large language model experiments.

\bibliography{refs}

\begin{thebibliography}{27}
\providecommand{\natexlab}[1]{#1}
\providecommand{\url}[1]{\texttt{#1}}
\expandafter\ifx\csname urlstyle\endcsname\relax
  \providecommand{\doi}[1]{doi: #1}\else
  \providecommand{\doi}{doi: \begingroup \urlstyle{rm}\Url}\fi

\bibitem[Ad{\'a}mek et~al.(1990)Ad{\'a}mek, Herrlich, and Strecker]{AHS1990}
Ji{\v{r}}{\'\i} Ad{\'a}mek, Horst Herrlich, and George Strecker.
\newblock \emph{Abstract and concrete categories}.
\newblock Wiley-Interscience, 1990.

\bibitem[Ad{\'a}mek et~al.(2021)Ad{\'a}mek, Chen, Milius, and Urbat]{ACMU2021}
Ji{\v{r}}{\'\i} Ad{\'a}mek, Liang-Ting Chen, Stefan Milius, and Henning Urbat.
\newblock Reiterman’s theorem on finite algebras for a monad.
\newblock \emph{ACM Transactions on Computational Logic (TOCL)}, 22\penalty0
  (4):\penalty0 1--48, 2021.

\bibitem[{Anthropic}(2025)]{ClaudeOpus4.5}
{Anthropic}.
\newblock System card: Claude opus 4.5.
\newblock Anthropic, November 2025.
\newblock URL \url{https://www.anthropic.com/claude-opus-4-5-system-card}.
\newblock Released November 24, 2025. Accessed on \today.

\bibitem[Asher(2025)]{Asher_LeanExplore_2025}
Justin Asher.
\newblock {LeanExplore: A search engine for Lean 4 declarations}, 2025.
\newblock URL \url{https://arxiv.org/abs/2506.11085}.

\bibitem[Azerbayev et~al.(2022)Azerbayev, Piotrowski, and Avigad]{proofNet}
Zhangir Azerbayev, Bartosz Piotrowski, and Jeremy Avigad.
\newblock {ProofNet}: A benchmark for autoformalizing and formally proving
  undergraduate-level mathematics problems.
\newblock In \emph{Second MATH-AI Workshop}, 2022.

\bibitem[Chen et~al.(2025)Chen, Chen, Du, Hu, Jiang, Jie, Jin, Jin, Li, Shi,
  et~al.]{SeepProver1.5}
Jiangjie Chen, Wenxiang Chen, Jiacheng Du, Jinyi Hu, Zhicheng Jiang, Allan Jie,
  Xiaoran Jin, Xing Jin, Chenggang Li, Wenlei Shi, et~al.
\newblock Seed-prover 1.5: Mastering undergraduate-level theorem proving via
  learning from experience.
\newblock \emph{arXiv preprint arXiv:2512.17260}, 2025.

\bibitem[Chen et~al.(2021)Chen, Tworek, Jun, Yuan, Pinto, Kaplan, Edwards,
  Burda, Joseph, Brockman, et~al.]{pass-k}
Mark Chen, Jerry Tworek, Heewoo Jun, Qiming Yuan, Henrique Ponde De~Oliveira
  Pinto, Jared Kaplan, Harri Edwards, Yuri Burda, Nicholas Joseph, Greg
  Brockman, et~al.
\newblock Evaluating large language models trained on code.
\newblock \emph{arXiv preprint arXiv:2107.03374}, 2021.

\bibitem[Chen(2021)]{Chen2021}
Ruiyuan Chen.
\newblock On sifted colimits in the presence of pullbacks.
\newblock \emph{arXiv preprint arXiv:2109.12708}, 2021.

\bibitem[{Gemini Team, Google}(2025)]{Gemini3pro}
{Gemini Team, Google}.
\newblock Gemini 3 pro model card.
\newblock November 2025.
\newblock URL
  \url{https://storage.googleapis.com/deepmind-media/Model-Cards/Gemini-3-Pro-Model-Card.pdf}.

\bibitem[Hubert et~al.(2025)Hubert, Mehta, Sartran, Horv{\'a}th,
  {\v{Z}}u{\v{z}}i{\'c}, Wieser, Huang, Schrittwieser, Schroecker, Masoom,
  et~al.]{DeepMind2025}
Thomas Hubert, Rishi Mehta, Laurent Sartran, Mikl{\'o}s~Z Horv{\'a}th, Goran
  {\v{Z}}u{\v{z}}i{\'c}, Eric Wieser, Aja Huang, Julian Schrittwieser, Yannick
  Schroecker, Hussain Masoom, et~al.
\newblock Olympiad-level formal mathematical reasoning with reinforcement
  learning.
\newblock \emph{Nature}, pp.\  1--3, 2025.

\bibitem[Jiang et~al.(2025)Jiang, He, Wang, Gao, Hu, Wang, Guan, Wu, Dai, Xiao,
  et~al.]{Fate}
Jiedong Jiang, Wanyi He, Yuefeng Wang, Guoxiong Gao, Yongle Hu, Jingting Wang,
  Nailing Guan, Peihao Wu, Chunbo Dai, Liang Xiao, et~al.
\newblock Fate: A formal benchmark series for frontier algebra of multiple
  difficulty levels.
\newblock \emph{arXiv preprint arXiv:2511.02872}, 2025.

\bibitem[Kong()]{Kong}
Liang Kong.
\newblock Category theory.
\newblock Unpublic.

\bibitem[Lin et~al.(2025)Lin, Tang, Lyu, Yang, Chung, Zhao, Jiang, Geng, Ge,
  Sun, et~al.]{goedel}
Yong Lin, Shange Tang, Bohan Lyu, Ziran Yang, Jui-Hui Chung, Haoyu Zhao, Lai
  Jiang, Yihan Geng, Jiawei Ge, Jingruo Sun, et~al.
\newblock Goedel-prover-v2: Scaling formal theorem proving with scaffolded data
  synthesis and self-correction.
\newblock \emph{arXiv preprint arXiv:2508.03613}, 2025.

\bibitem[Liu et~al.(2025)Liu, Mei, Lin, Xue, Wang, Xu, Wu, Zhang, Lin, Dong,
  et~al.]{DeepseekV3.2}
Aixin Liu, Aoxue Mei, Bangcai Lin, Bing Xue, Bingxuan Wang, Bingzheng Xu,
  Bochao Wu, Bowei Zhang, Chaofan Lin, Chen Dong, et~al.
\newblock Deepseek-v3. 2: Pushing the frontier of open large language models.
\newblock \emph{arXiv preprint arXiv:2512.02556}, 2025.

\bibitem[Liu et~al.(2023)Liu, Shen, Xin, Liu, Yuan, Wang, Ju, Zheng, Yin, Li,
  et~al.]{FIMO}
Chengwu Liu, Jianhao Shen, Huajian Xin, Zhengying Liu, Ye~Yuan, Haiming Wang,
  Wei Ju, Chuanyang Zheng, Yichun Yin, Lin Li, et~al.
\newblock {FIMO}: A challenge formal dataset for automated theorem proving.
\newblock \emph{arXiv preprint arXiv:2309.04295}, 2023.

\bibitem[Mac~Lane(1998)]{ML1998}
Saunders Mac~Lane.
\newblock \emph{Categories for the working mathematician}, volume~5.
\newblock Springer Science \& Business Media, 1998.

\bibitem[mathlib Community(2020)]{mathlib}
The mathlib Community.
\newblock The {Lean} mathematical library.
\newblock In \emph{Proceedings of the 9th ACM SIGPLAN International Conference
  on Certified Programs and Proofs}, 2020.

\bibitem[{OpenAI}(2025)]{ChatGPT5.2}
{OpenAI}.
\newblock Update to gpt-5 system card: Gpt-5.2.
\newblock OpenAI, December 2025.
\newblock URL \url{https://openai.com/index/gpt-5-system-card-update-gpt-5-2/}.
\newblock Released December 11, 2025. Accessed on \today.

\bibitem[Polu et~al.(2022)Polu, Han, Zheng, Baksys, Babuschkin, and
  Sutskever]{Openai2022}
Stanislas Polu, Jesse~Michael Han, Kunhao Zheng, Mantas Baksys, Igor
  Babuschkin, and Ilya Sutskever.
\newblock Formal mathematics statement curriculum learning.
\newblock \emph{arXiv preprint arXiv:2202.01344}, 2022.

\bibitem[Ren et~al.(2025)Ren, Shao, Song, Xin, Wang, Zhao, Zhang, Fu, Zhu,
  Yang, et~al.]{deepseekproverv2}
ZZ~Ren, Zhihong Shao, Junxiao Song, Huajian Xin, Haocheng Wang, Wanjia Zhao,
  Liyue Zhang, Zhe Fu, Qihao Zhu, Dejian Yang, et~al.
\newblock {D}eep{S}eek-{P}rover-{V}2: Advancing formal mathematical reasoning
  via reinforcement learning for subgoal decomposition.
\newblock \emph{arXiv preprint arXiv:2504.21801}, 2025.

\bibitem[Riehl(2017)]{Riehl2017}
Emily Riehl.
\newblock \emph{Category theory in context}.
\newblock Courier Dover Publications, 2017.

\bibitem[Shang et~al.(2025)Shang, Wan, Peng, Wu, hui Chen, Yan, and
  Zhang]{stepfunprover2025}
Shijie Shang, Ruosi Wan, Yue Peng, Yutong Wu, Xiong hui Chen, Jie Yan, and
  Xiangyu Zhang.
\newblock Stepfun-prover preview: Let's think and verify step by step, 2025.
\newblock URL \url{https://arxiv.org/abs/2507.20199}.

\bibitem[Team et~al.(2025)Team, Bai, Bao, Chen, Chen, Chen, Chen, Chen, Chen,
  Chen, et~al.]{KimiK2}
Kimi Team, Yifan Bai, Yiping Bao, Guanduo Chen, Jiahao Chen, Ningxin Chen,
  Ruijue Chen, Yanru Chen, Yuankun Chen, Yutian Chen, et~al.
\newblock Kimi k2: Open agentic intelligence.
\newblock \emph{arXiv preprint arXiv:2507.20534}, 2025.

\bibitem[Tsoukalas et~al.(2024)Tsoukalas, Lee, Jennings, Xin, Ding, Jennings,
  Thakur, and Chaudhuri]{putnamBench}
George Tsoukalas, Jasper Lee, John Jennings, Jimmy Xin, Michelle Ding, Michael
  Jennings, Amitayush Thakur, and Swarat Chaudhuri.
\newblock {P}utnam{B}ench: Evaluating neural theorem-provers on the putnam
  mathematical competition.
\newblock In A.~Globerson, L.~Mackey, D.~Belgrave, A.~Fan, U.~Paquet,
  J.~Tomczak, and C.~Zhang (eds.), \emph{Advances in Neural Information
  Processing Systems}, volume~37, pp.\  11545--11569. Curran Associates, Inc.,
  2024.
\newblock URL
  \url{https://proceedings.neurips.cc/paper_files/paper/2024/file/1582eaf9e0cf349e1e5a6ee453100aa1-Paper-Datasets_and_Benchmarks_Track.pdf}.

\bibitem[Wang et~al.(2025)Wang, Unsal, Lin, Baksys, Liu, Santos, Sung, Vinyes,
  Ying, Zhu, et~al.]{kimina}
Haiming Wang, Mert Unsal, Xiaohan Lin, Mantas Baksys, Junqi Liu, Marco~Dos
  Santos, Flood Sung, Marina Vinyes, Zhenzhe Ying, Zekai Zhu, et~al.
\newblock Kimina-prover preview: Towards large formal reasoning models with
  reinforcement learning.
\newblock \emph{arXiv preprint arXiv:2504.11354}, 2025.

\bibitem[Zheng()]{Zheng}
Hao Zheng.
\newblock Category note.
\newblock Unpublic.

\bibitem[Zheng et~al.(2022)Zheng, Han, and Polu]{miniF2F}
Kunhao Zheng, Jesse~Michael Han, and Stanislas Polu.
\newblock {MiniF2F}: A cross-system benchmark for formal olympiad-level
  mathematics.
\newblock In \emph{The Tenth International Conference on Learning
  Representations}, 2022.

\end{thebibliography}
\bibliographystyle{refs}

\newpage
\appendix

\section{Detailed Benchmark Breakdown}\label{app:topic_details}
In this section, we provide a detailed description of the eight topical clusters comprising LeanCat, complementing the overview in Section \ref{sec:methodology}.

\begin{itemize}
    \item \textbf{Basic Category Properties (Problems 1-18)}: Fundamental results about categories and morphisms. These include properties of monomorphisms and epimorphisms, initial/terminal objects, factorization of idempotents, and examples of categorical constructions.
    \item \textbf{Adjunctions (Problems 19-29)}: Constructions and criteria involving adjoint functors, a central concept in category theory. Problems include proving that familiar functors have left or right adjoints, as well as general adjointness criteria like the comma category condition (Problem 19) and some concrete examples (Problem 28). These tasks test a prover's ability to manipulate universal properties and to move between element-wise (“pointwise”) reasoning and diagrammatic reasoning.
    \item \textbf{Reflective and Coreflective Subcategories (Problems 30-33)}: Abstract properties and concrete examples of a special kind of subcategories (e.g. classifying the reflective subcategories of $\mathcal{S}et$ and $\mathcal{T}\mathrm{op}^{CH}$).
    \item \textbf{Concrete Categories (Problems 34-41)}: Categories with a faithful forgetful functor to Set and related notions. These tasks are largely concrete and overlap substantially with other areas of mathematics, such as topology, order theory, and set theory. They are designed to test a model's ability to connect abstract concepts with concrete examples.
    \item \textbf{Limits and Colimits (Problems 42-73)}: This is the largest cluster, covering a range of results about limits, colimits, and related categorical constructs. 
    Many of these statements are at the frontier of what Lean's Mathlib currently contains, indeed some (like Problem 46 or 67) required developing new formal definitions. 
    This cluster stresses a prover's ability to chain together multiple categorical facts.
    \item \textbf{Cocompletions (Problems 74-78)}: This part is based on recent work on cocompletions. It requires the LLM to introduce new definitions and then prove the key theorems built on those definitions that are not currently available in Mathlib.
    \item \textbf{Abelian Categories (Problems 79-90)}: Tasks concerning abelian categories and homological algebra concepts. Abelian categories are highly structured categories (every morphism has a kernel and cokernel, etc.) and generalize the category of modules or abelian groups. 
    These statements mirror standard results in homological algebra, but formalizing them in Lean requires care with categorical abstractions (e.g. kernel objects, exact sequences) that are more complex than their set-based counterparts. 
    Solving them may require the prover to introduce creative auxiliary lemmas about kernels, images, or exactness - a tall order for automated tools.
    \item \textbf{Monads (Problems 91-100)}: The final cluster centers on monads and their associated constructions (Kleisli and Eilenberg-Moore categories). 
    Monads are a high-level concept that encapsulates a form of “computation” or structure; proving properties about them in Lean often demands a two-level reasoning (reasoning about the monad's algebraic laws and about category-theoretic conditions like preservation of coequalizers). 
    This cluster provides a valuable test of an AI's ability to work with highly abstract algebraic structure in a category-theoretic setting.
\end{itemize}

\subsection{Sample Problem Instance}
To illustrate the data format of LeanCat, we present two sample entries, each problem consists of a natural language description (stored in a companion Markdown file) and a self-contained Lean 4 file with the formal declaration.

\begin{figure}[h]
    \centering
    \begin{minipage}{0.9\textwidth}
    
    % --- 1. Natural Language Part ---
    \hrule height 1pt
    \vspace{0.5em}
    \textbf{\textsf{Natural Language Statement (.md)}}\\
    \vspace{0.5em}
    \textit{Definition:} A functor $F : \mathcal C \to \mathcal D$ is said to $\textbf{lift limits}$ if for every diagram $D: \mathcal I\to \mathcal C$ and every limit $L$ of $F\circ D$, there exists a limit $L'\in\mathcal D$ such that $F(L')\cong L$.

    \textit{Theorem:} There is a functor that lifts limits but is not faithful.
    
    \vspace{0.8em}
    \hrule
    \vspace{0.8em}
    
    % --- 2. Formal Statement Part ---
    \textbf{\textsf{Formal Statement (.lean)}}\\
    % \vspace{0.0em}
    \includegraphics[width=0.8\textwidth]{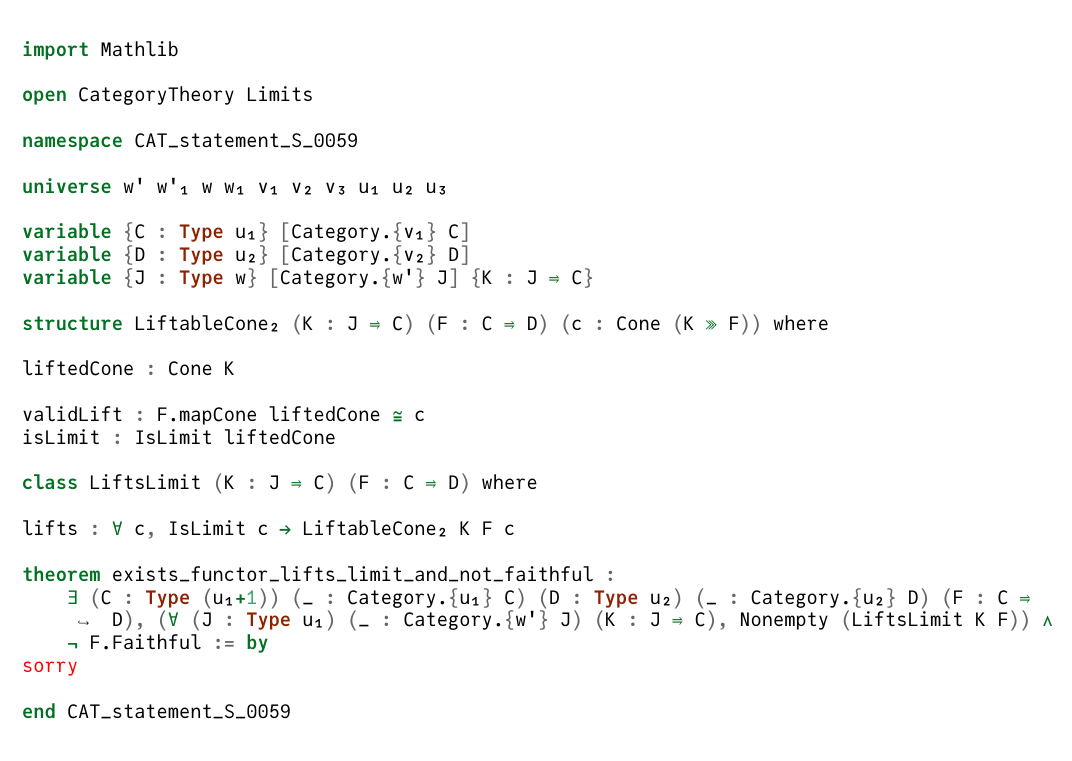}  
    \vspace{0.2em}
    \hrule height 1pt
    \end{minipage}
    \caption{\textbf{A Sample LeanCat Task (Problem 59).} The task is to prove the existence of a functor that lifts limits but is not faithful.}
    \label{fig:sample_task}
\end{figure}

\begin{figure}[h]
    \centering
    \begin{minipage}{0.9\textwidth}
    
    % --- 1. Natural Language Part ---
    \hrule height 1pt
    \vspace{0.5em}
    \textbf{\textsf{Natural Language Statement (.md)}}\\
    \vspace{0.5em}
    \textit{Definition:} Let $\mathcal C$ be a locally small category. An object $c \in\mathcal C$ is called {\bf compact} if $\mathrm{hom}_{\mathcal C} (c,-)$ preserves filtered colimits.

    \textit{Theorem:} In the category $\mathcal{G}\mathrm{rp}$ of groups, an object is compact if and only if it is a finitely presented group. Every group is a filtered colimit of finitely presented groups.
    
    \vspace{0.8em}
    \hrule
    \vspace{0.8em}
    
    % --- 2. Formal Statement Part ---
    \textbf{\textsf{Formal Statement (.lean)}}\\
    % \vspace{0.0em}
    \includegraphics[width=0.8\textwidth]{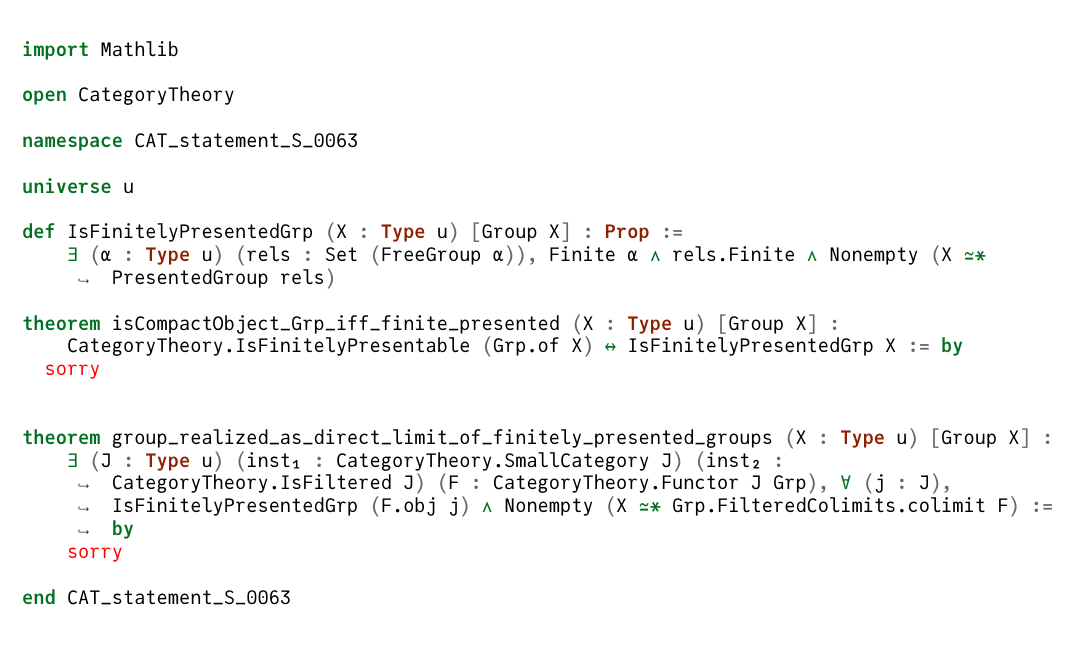}  
    \vspace{0.2em}
    \hrule height 1pt
    \end{minipage}
    \caption{\textbf{A Sample LeanCat Task (Problem 63).} The task is to prove the compact object in $\mathcal G\mathrm{rp}$ is exactly the finitely presented groups.}
    \label{fig:prb_63}
\end{figure}

\section{Difficulty Scoring Data}
We provide the full raw data used for difficulty scoring:

\begin{figure}[h]
    \centering
    \includegraphics[width=0.9\textwidth]{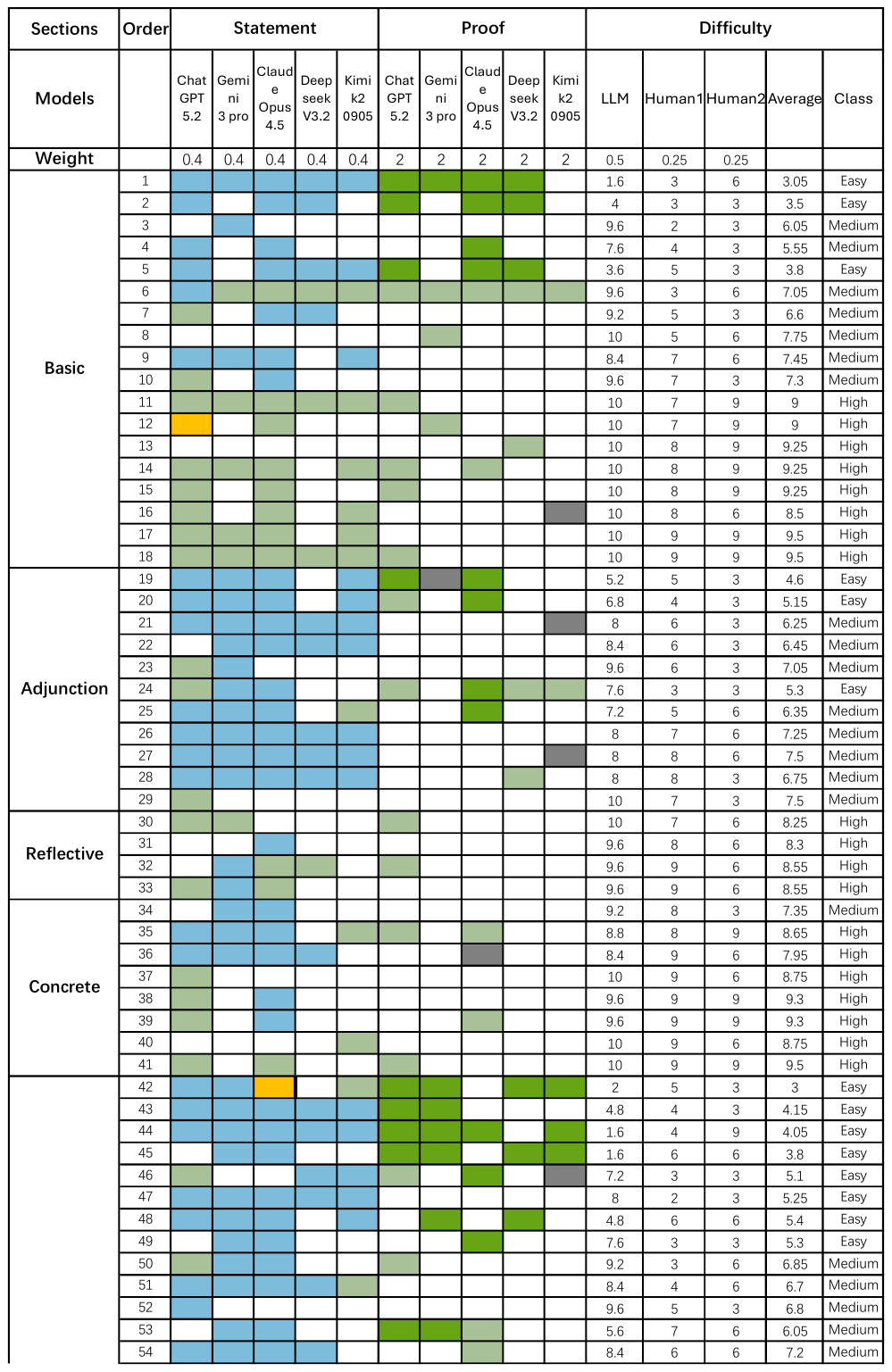}
    \caption{Snapshot of the difficulty annotation data. \textbf{Green}: Correct Lean proof. \textbf{Blue}: Correct formal statement. \textbf{Olive}: Statement correct but language mismatch. \textbf{Grey}: Syntax/Type error (incomplete Lean script). \textbf{Yellow}: Mathematically correct but informal Lean code.}
    \label{app:fig:score_sample}
\end{figure}

\begin{figure}[h]
    \centering
    \includegraphics[width=0.9\textwidth]{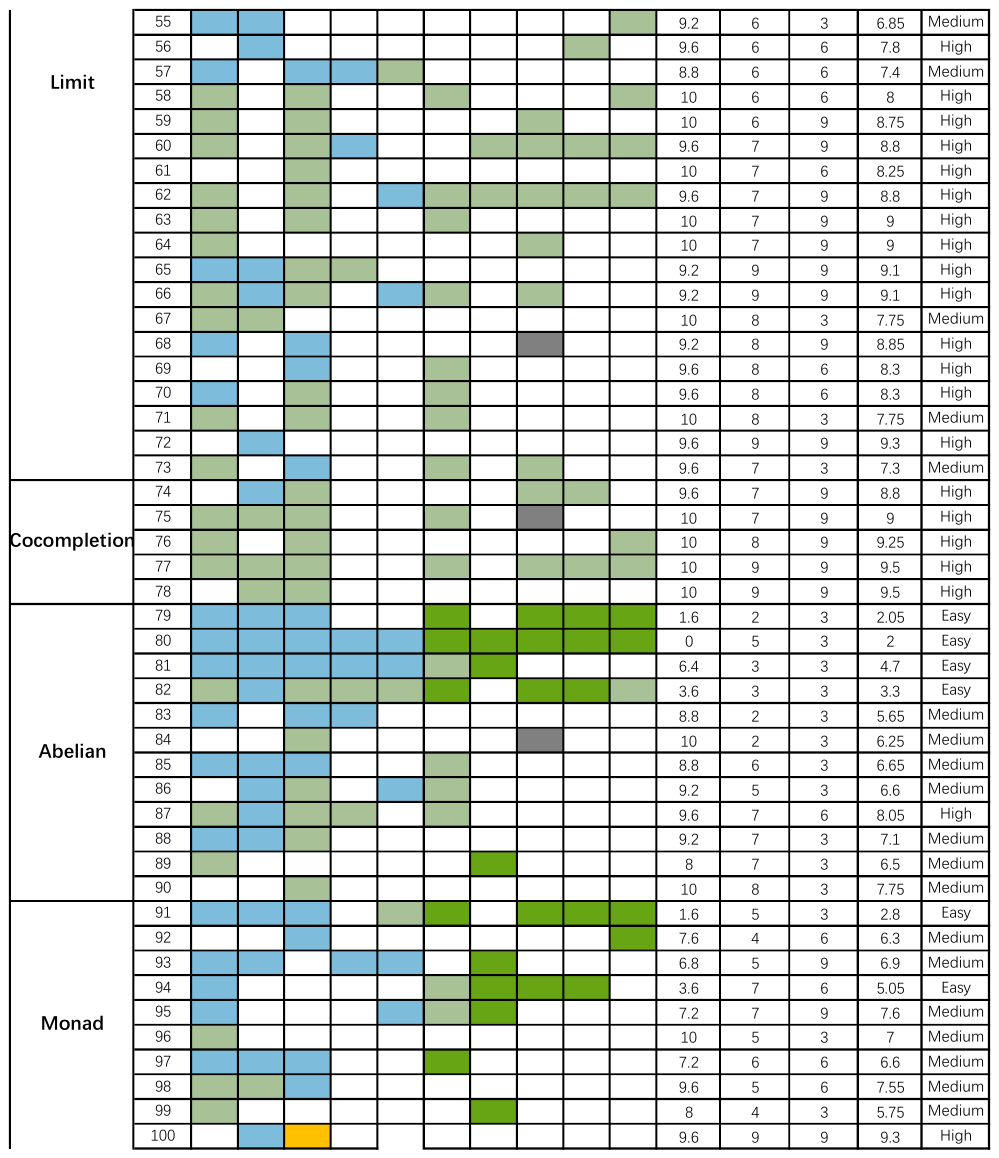}
    \caption{Snapshot of the difficulty annotation data (continued). \textbf{Green}: Correct Lean proof. \textbf{Blue}: Correct formal statement. \textbf{Olive}: Statement correct but language mismatch. \textbf{Grey}: Syntax/Type error (incomplete Lean script). \textbf{Yellow}: Mathematically correct but informal Lean code.}
    \label{app:fig:score_sample_ctd}
\end{figure}

\clearpage
\section{Visualization of Model Performance}\label{app:visualize}
We provide a detailed visualization of the impact of formal statement guidance across all models:
\begin{figure}[h]
    \centering
    \includegraphics[width=\textwidth]{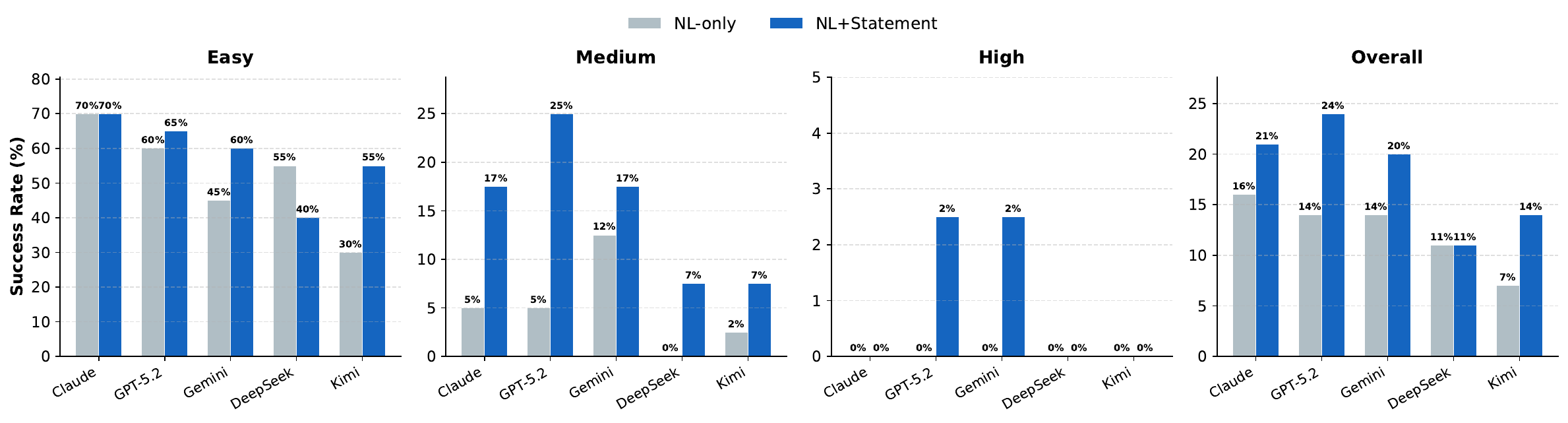}
    \caption{\textbf{Impact of Formal Statement Guidance across All Models (Ablation Study).} 
    Comparison of LeanBridge performance with \textit{NL-only} (grey) vs. \textit{NL+Statement} (blue) inputs. 
    Providing formal statements consistently improves success rates on \textbf{Medium} tasks for nearly all models. Notably, it enables GPT-5.2 and Gemini-3-Pro to solve \textbf{High} difficulty problems (the 2.5\% breakthrough), which are otherwise unsolvable.}
    \label{app:fig:ablation_detailed}
\end{figure}

\section{Prompts and Experimental Setup}\label{app:prompts}

\subsection{System Prompts}
We use the following system prompt for the baseline models (pass@k):

\begin{lstlisting}[
    language=lean,
    caption={Prompt template for pass@k proof generation},
    label={lst:prompt}     
]
    """
    You are an expert in Lean 4 and Mathematics. Please finish the following proof in Lean4 code. 
    Do not change the original statement. Copy the final statement to prove exactly. 
    Please include the complete header (including imports and namespaces) so that your code can 
    pass the Lean4 compiler. Please solve the statement step by step and provide your complete 
    Lean4 code between ```lean4 and ``` after careful reasoning. 
    The statement for you to complete is: 

    ```lean4
        {formal_statement}
    ```
    """
\end{lstlisting}

\subsection{LeanBridge Prompts}
For the LeanBridge agent, we use a distinct prompt structure to facilitate the \emph{retrieve-generate-verify} loop:
\begin{lstlisting}[
    language=lean,
    caption={Prompt template for LeanBridge proof generation},
    label={lst:leanbridge_proof_prompt}     
]
    """You are an expert in Lean 4 and Mathematics. Please finish the following proof in Lean4 code. 

    if provided a formal statement, do not change the original formal statement. Copy the final statement to prove exactly. Please include the complete header (including imports and namespaces) so that your code can pass the Lean4 compiler. 
    
    Please solve the statement step by step and provide your complete Lean4 code between ```lean4 and ``` after careful reasoning.
    
    The statement for you to complete is: 
    Natural language statement:
    {nl_statement}

    Lean 4 formal statement:
    {formal_statement}
    
    Reference Knowledge (Search Results from Mathlib):
    {search_contents}

    """
\end{lstlisting}

\begin{lstlisting}[
    language=lean,
    caption={Prompt template for LeanBridge refinement},
    label={lst:leanbridge_refine_prompt}     
]
    """The previous Lean 4 proof attempt contains errors. Your goal is to fix them according to the compiler's error message.

    Input:
    Lean 4 code:
    {original_proof}

    Error messages:
    {error_messages}

    Instructions:
    1. First, analyze the error message carefully in a "## Error Analysis" section. Identify specifically why the verification failed (e.g., missing import, wrong tactic, type mismatch, or a missing lemma).
    2. Based on your analysis, decide whether you need to search Mathlib for external knowledge (definitions, theorems, or syntax) or if you can fix it immediately.

    Condition A: If you need external information
    Output the search query wrapped in the tag below. Make your query specific (e.g., a math statement, a definition name, or a type signature).
    Format:
    ## Error Analysis
    (Your analysis of the error and what you need to find)
    [SEARCH: your_search_query]

    Condition B: If you can fix it immediately
    Output the analysis and the fully refined code.
    Format:
    ## Error Analysis
    (Your analysis of the error and the fix plan)

    ## Refined Code
    ```lean
    -- The complete, corrected Lean 4 code
    ```
    """
\end{lstlisting}

\subsection{Hyperparameters}
\begin{table}[h]
    \centering
    \caption{Detailed experimental configuration and hyperparameters.}
    \label{tab:exp_setup_details}
    \begin{tabular}{l l p{6cm}}
    \toprule
    \textbf{Component} & \textbf{Parameter} & \textbf{Value / Description} \\
    \midrule
    \textbf{LLM Generation} & Temperature & $1.0$ (Encouraging diversity) \\
                            & Max Tokens & $50,000$ (Generalist models) \\
                            &            & \textit{Model-specific max} (Specialized provers) \\
    \midrule
    \textbf{Verification}   & Lean Version & v4.19.0 \\
                            & Timeout & $300$ seconds per attempt \\
    \midrule
    \textbf{LeanBridge}     & Loop Limit & Max $4$ refinement iterations \\
                            & Retrieval & Top-$3$ results per search query \\
    \bottomrule
    \end{tabular}
\end{table}

\section{Case Study}\label{app:case_study}

This section presents a detailed case study of a specific problem, including the original problem statement, the generated Lean 4 code, and the evaluation of the generated code.

\subsection{Success Case}
\begin{figure}[h]
    \centering
    \begin{minipage}{0.9\textwidth}
    
    % --- 1. Natural Language Part ---
    \hrule height 1pt
    \vspace{0.5em}    
    % --- 2. Formal Statement Part ---
    \textbf{\textsf{Problem 59 Proving Trajectory}}\\
    % \vspace{0.0em}
    \includegraphics[width=0.8\textwidth]{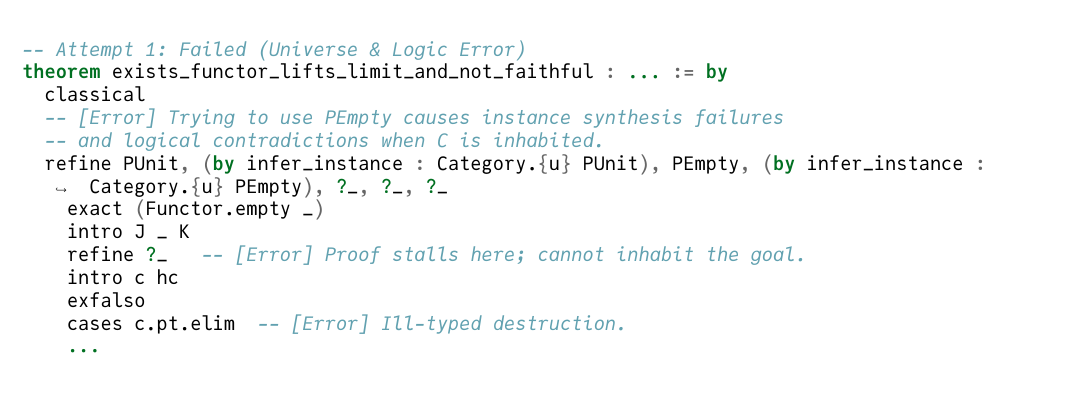}  
    \vspace{0.2em}
    \hrule

    \includegraphics[width=0.8\textwidth]{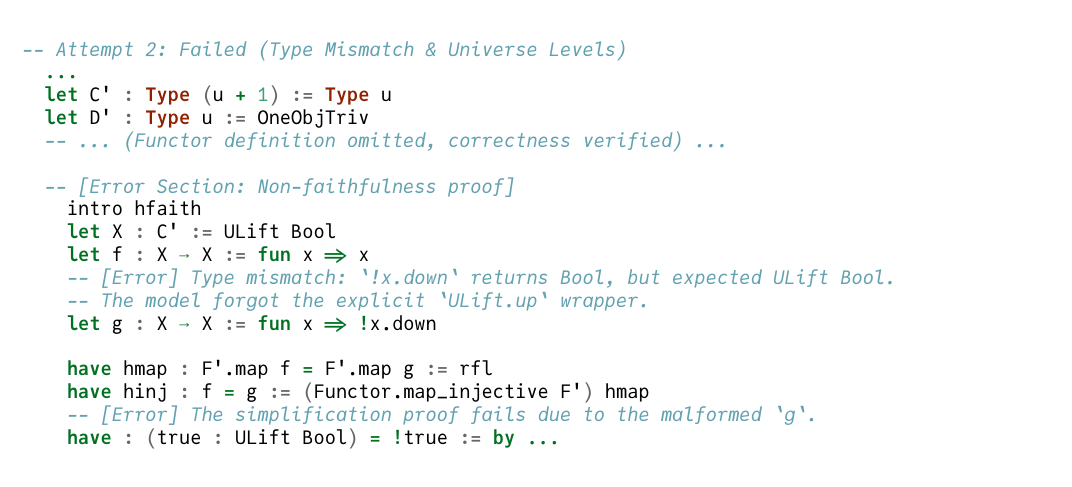} 
    \vspace{0.2em}
    \hrule
    \includegraphics[width=0.8\textwidth]{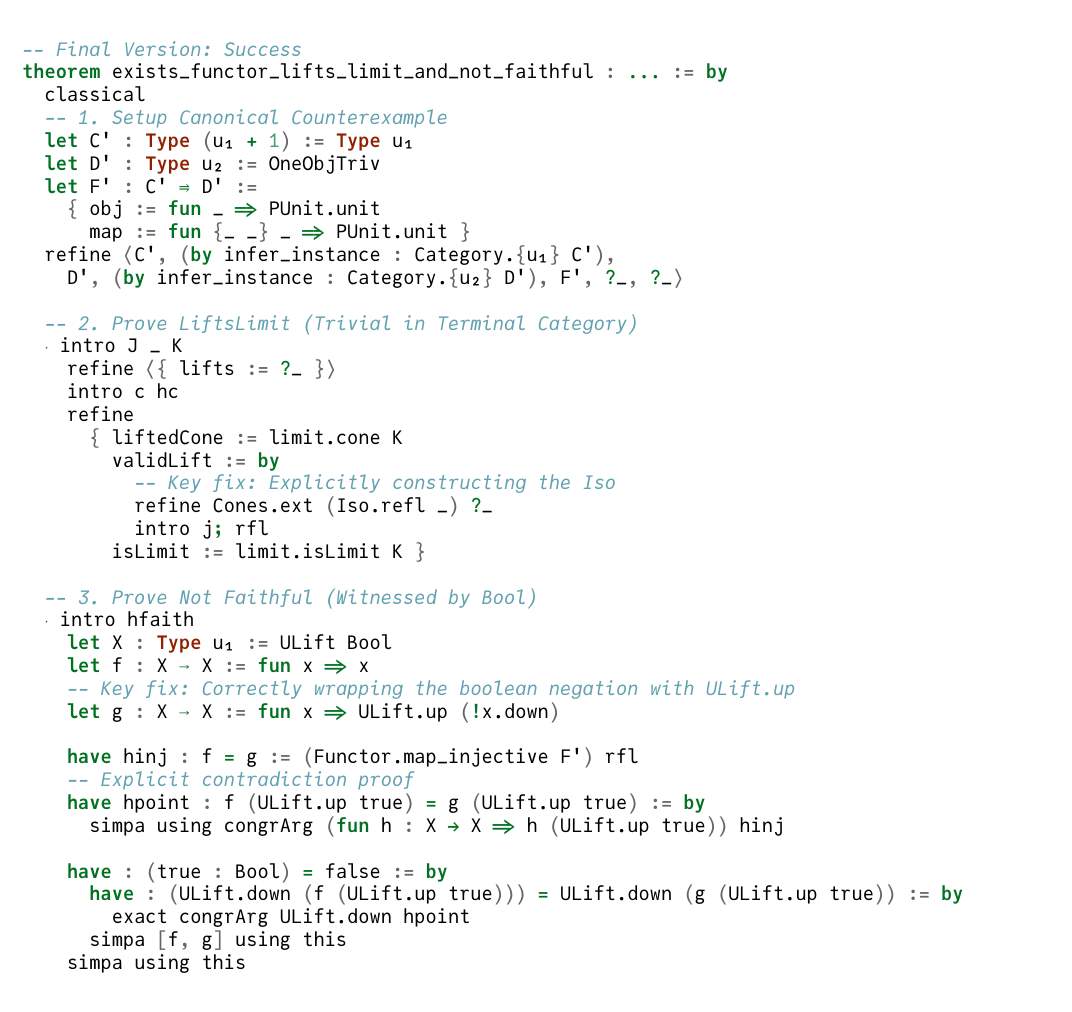} 
    \vspace{0.2em}
    \hrule height 1pt
    \end{minipage}
    \caption{\textbf{Case Study (Success): Iterative Refinement on Problem 59.} The proving trajectory by LeanBridge (GPT-5.2), showing the progression from initial attempts to a successful formal proof. }
    \label{fig:case_study_59}
\end{figure}

As depicted in Appendix Figure \ref{fig:sample_task}, Problem 59 is a representative success case: it is the only \textbf{High}-difficulty problem solved in our setting, achieved only by GPT-5.2 and Gemini-3-Pro with LeanBridge.
We study the GPT-5.2 attempt.
This run fails in early attempts but verifies successfully on the final refinement (the proving trajectory showed in Figure \ref{fig:case_study_59}).

\paragraph{1. Strategy used (and mathematical correctness).}
The final successful witness takes
\[
C := \mathrm{Type}\,u_1,\qquad
D := \text{a terminal category on } \texttt{PUnit},\qquad
F : C \to D \text{ the constant functor}.
\]
This strategy is mathematically appropriate: limit-lifting becomes easy because $C$ has all limits and $D$ has trivial cone structure, while non-faithfulness is immediate because $F$ collapses distinct morphisms to the unique morphism in $D$.

\paragraph{2. Why it fails initially, and how LLM fixes it.}
The first attempt tries to exploit ``vacuity'' by taking an empty codomain (e.g.\ \texttt{PEmpty}) and concludes lifting should be trivial. This fails due to (i) universe/typeclass issues when synthesizing the required \texttt{Category} instances and (ii) ill-typed constructions (e.g.\ attempting to inhabit \texttt{PEmpty}).
In subsequent iterations, the run pivots to the clean terminal-codomain construction above, but still encounters Lean-engineering obstacles:
universe mismatches when using \texttt{Bool : Type 0} as an object of \texttt{Type u\_1}, and brittle pattern matching tactics (e.g.\ \texttt{casesm*}).
The final repair commits to the clean construction and patches these issues by using \texttt{ULift Bool : Type u\_1} (with explicit \texttt{ULift.up}) and replacing brittle matches with simple \texttt{cases}.

\paragraph{3. Why the fix works.}
Limit lifting succeeds because \texttt{Type u\_1} already provides canonical limit data \texttt{limit.cone K} and \texttt{limit.isLimit K}, which can be used verbatim as the lift.
The required cone isomorphism is trivial in a terminal category: all cone legs are definitional equal to the unique morphism, so the isomorphism can be constructed by extensionality with reflexive components.
Non-faithfulness is witnessed concretely by two distinct endomorphisms on \texttt{ULift Bool} (identity vs.\ boolean negation), which both map to the unique morphism in $D$; this contradicts \texttt{Functor.map\_injective} by evaluating at \texttt{ULift.up true}.

In summary, the run succeeds because it converges to a canonical, Lean-friendly counterexample: a constant functor from a complete category (\texttt{Type}) to a terminal category.
This choice simultaneously trivializes the lifting obligation (via existing limit infrastructure in \texttt{Type}) and makes non-faithfulness explicit (via collapsing distinct functions), while the final patches resolve the remaining universe and pattern-matching pitfalls.

\subsection{Failure Case}
\begin{figure}[h]
    \centering
    \begin{minipage}{0.9\textwidth}
    
    % --- 1. Natural Language Part ---
    \hrule height 1pt
    \vspace{0.5em}    
    % --- 2. Formal Statement Part ---
    \textbf{\textsf{Problem 63 Proving Trajectory}}\\
    % \vspace{0.0em}
    \includegraphics[width=0.8\textwidth]{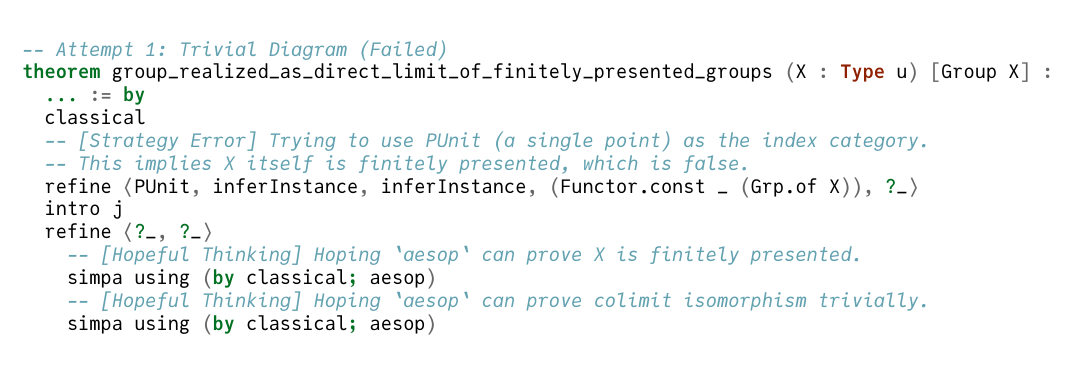}  
    \vspace{0.2em}
    \hrule

    \includegraphics[width=0.8\textwidth]{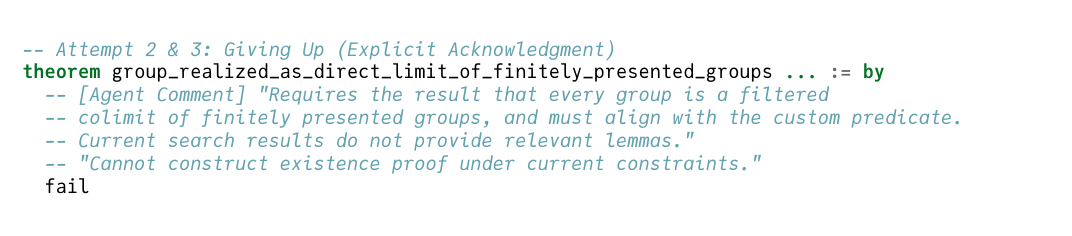} 
    \vspace{0.2em}
    \hrule
    \includegraphics[width=0.8\textwidth]{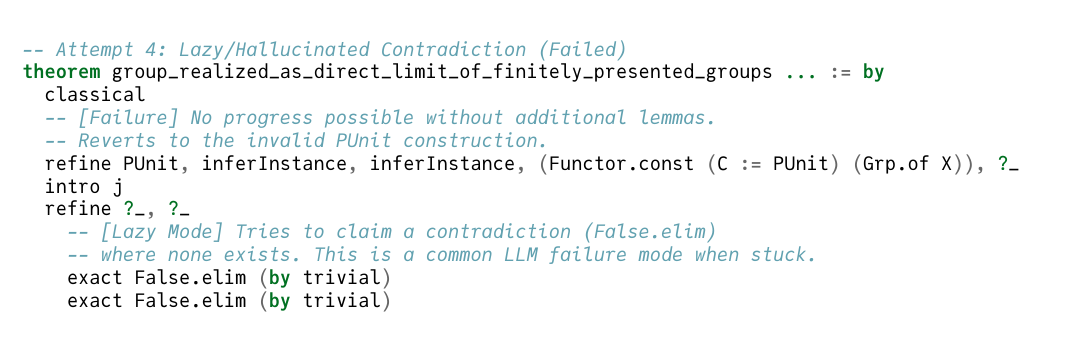} 
    \vspace{0.2em}
    \hrule height 1pt
    \end{minipage}
    \caption{\textbf{Case Study (Failure): Iterative Refinement on Problem 63.} The proving trajectory by LeanBridge (GPT-5.2).}
    \label{fig:case_study_63}
\end{figure}

We also study the GPT-5.2 with LeanBridge.
As depicted in Appendix Figure \ref{fig:prb_63}, Problem 63 is a representative failure case (the proving trajectory showed in Figure \ref{fig:case_study_63}): combining both a mathematical-strategy error and a “lazy” failure mode.

The task is to formalize in Lean + mathlib the claim that in $\mathcal{G}\mathrm{rp}$, categorical finite presentability matches finite group presentation (generators/relations), and that every group is a filtered colimit of finitely presented groups. In the file, ``finitely presented as a group'' is encoded by a custom predicate
\[
\begin{aligned}
\texttt{IsFinitelyPresentedGrp } X :={}& \exists (\alpha, \mathrm{rels}), \texttt{Finite } \alpha \ \wedge \\
& \texttt{rels.Finite } \wedge \texttt{Nonempty} (X \simeq_* \texttt{PresentedGroup rels})
\end{aligned}
\]
and the goals are (i) an equivalence
\[
\texttt{IsFinitelyPresentable (Grp.of X)} \;\leftrightarrow\; \texttt{IsFinitelyPresentedGrp X},
\]
and (ii) a filtered diagram $F$ landing in groups satisfying \texttt{IsFinitelyPresentedGrp} whose colimit is isomorphic to $X$.

\paragraph{1. Strategy used (and mathematical correctness).}
The LLM aims to reuse existing mathlib results: reduce the statement to mathlib’s notion \texttt{IsFinitelyPresentable} in \texttt{Grp}, then connect that to the concrete \texttt{PresentedGroup}-based predicate, and finally invoke a filtered-colimit theorem in \texttt{Grp}. 
This high-level plan matches the correct mathematical strategy: one normally proves the equivalence via standard theorems (Baer/Freyd-style characterizations in algebraic categories) and then uses the classical fact that any group is a filtered colimit of finitely presented groups. However, when it cannot find the required bridge theorems, the LLM falls back to an invalid shortcut: it tries to realize the filtered colimit using a trivial/constant diagram (e.g.\ indexed by \texttt{PUnit}), which would effectively force \texttt{IsFinitelyPresentedGrp X} for arbitrary $X$, a statement that is false in general.

\paragraph{2. Why it fails, and how the LLM tries to fix it.}
In the first attempt the LLM tries to close the equivalence using a guessed lemma name like \texttt{Grp.isFinitelyPresentable\_iff}, but Lean reports it as an unknown constant (the lemma does not exist under that name). 
For the colimit part it attempts to define a constant functor into a filtered index category and then discharge the remaining obligations with automation (e.g.\ \texttt{aesop}), but the hard goals remain: one must actually prove \texttt{IsFinitelyPresentedGrp (F.obj j)} for each index $j$ and produce \texttt{Nonempty (X} $\simeq_*$ \texttt{colimit F)}. 
After this failure the LLM “fixes” by switching placeholders: it inserts \texttt{fail} in place of the missing mathematics (which Lean rejects), then replaces \texttt{fail} by \texttt{simp} (which makes no progress on genuinely nontrivial goals), and finally tries contradiction-style discharge, attempting to reduce obligations to \texttt{False} and solve them via \texttt{assumption} or \texttt{rcases}. These last steps fail because no contradiction is derivable from the given hypotheses, and because some tactic steps are applied to ill-typed or non-inductive targets, indicating the repair loop has drifted away from the correct categorical structure.

\paragraph{3. Why the fixes do not work.}
The fixes fail for structural reasons rather than minor syntax. First, the proof fundamentally needs nontrivial ``bridge'' theorems (or a substantial new development) connecting mathlib’s categorical finite presentability in \texttt{Grp} with the specific user-defined predicate \texttt{IsFinitelyPresentedGrp} based on \texttt{PresentedGroup}. The LLM never finds or constructs this bridge; instead it guesses lemma names. Second, the constant-diagram shortcut cannot satisfy the statement ``every group is a filtered colimit of finitely presented groups,'' since it would imply every group is itself finitely presented. Third, swapping in \texttt{fail}, \texttt{simp}, or attempting to conjure \texttt{False} does not create the missing mathematical content; Lean correctly rejects these as non-proofs.

In summary, the run fails because the LLM does not supply the missing mathematical infrastructure required by the formal statement: (i) a correct, explicit connection between \texttt{IsFinitelyPresentable} in \texttt{Grp} and the concrete \texttt{PresentedGroup}-based predicate, and (ii) a correct nontrivial filtered diagram witnessing $X$ as a filtered colimit of finitely presented groups. When those are not available, the LLM falls back to lemma-name guessing and tactic placeholders, which do not typecheck or do not solve the goals, so verification never succeeds.

\section{Datasheet for Datasets}\label{app:datasheet}
\begin{itemize}
    \item \textbf{Motivation}: To benchmark library-level abstraction reasoning in LLMs.
    \item \textbf{Composition}: 100 pairs of Natural Language and Lean 4 statement code.
    \item \textbf{Collection Process}: Expert curation from textbooks, unpublished lecture note and research papers.
    \item \textbf{Maintenance}: Hosted on GitHub under MIT.
\end{itemize}

%%%%%%%%%%%%%%%%%%%%%%%%%%%%%%%%%%%%%%%%%%%%%%%%%%%%%%%%%%%%%%%%%%%%%%%%%%%%%%%
%%%%%%%%%%%%%%%%%%%%%%%%%%%%%%%%%%%%%%%%%%%%%%%%%%%%%%%%%%%%%%%%%%%%%%%%%%%%%%%

\end{document}